\def\fullversion{1}

\documentclass[envcountsame,envcountsect,orivec]{llncs}

\ifnum\fullversion=1\relax
  \usepackage[a4paper,margin=1.1in]{geometry}
\fi

\usepackage[T1]{fontenc}
\usepackage{microtype}

\usepackage{amsmath,amssymb,mathtools}

\usepackage{xcolor}
\usepackage{xurl}
\usepackage[normalem]{ulem}
\usepackage[
  colorlinks,
  allcolors=blue,
  bookmarksdepth=2
]{hyperref}

\ifnum\fullversion=1\relax
  \usepackage{etoolbox}
  \makeatletter
  \patchcmd{\thebibliography}
    {\small}
    {\def\@biblabel##1{[##1]}\small}
    {}
    {\PackageError{main}{Failed to set bracketed bibliography labels}{}}
  \makeatother
\fi

\newcommand{\Z}{\mathbb{Z}}
\newcommand{\Q}{\mathbb{Q}}
\newcommand{\R}{\mathbb{R}}
\newcommand{\C}{\mathbb{C}}
\newcommand{\ii}{\mathrm{i}}

\newcommand{\norm}[1]{\left\lVert#1\right\rVert}
\newcommand{\ip}[2]{\langle#1,#2\rangle}

\newcommand{\radius}{T}
\newcommand{\packbase}{\alpha}
\newcommand{\GU}{\mathcal{U}_m}

\newcommand{\SSum}{\textup{Subset Sum}}
\newcommand{\NP}{\mathsf{NP}}
\newcommand{\SVP}{\mathsf{SVP}}
\newcommand{\GapSVP}{\mathsf{GapSVP}}
\newcommand{\ISVP}{\mathsf{IdealSVP}}
\newcommand{\MSVP}{\mathsf{ModuleSVP}}

\newcommand{\polyd}{d}
\newcommand{\polyk}{k}
\newcommand{\polyae}{{a_E}}
\newcommand{\polyt}{t}
\newcommand{\polytzero}{{t_0}}
\newcommand{\polyres}{v}
\newcommand{\polyu}{u}
\newcommand{\polyh}{h}
\newcommand{\polyhbase}{\tilde{h}}
\newcommand{\polyv}{v}
\newcommand{\polyq}{q}
\newcommand{\polyazero}{{p^{(0)}}}
\newcommand{\polyaone}{{p^{(1)}}}
\newcommand{\polyg}{g}

\newcommand{\matG}{\mathbf{G}}
\newcommand{\matM}{\mathbf{M}}
\newcommand{\matB}{\mathbf{B}}
\newcommand{\matI}{\mathbf{I}}

\newcommand{\vecxi}{\boldsymbol{\xi}}
\newcommand{\vecy}{\mathbf{y}}
\newcommand{\vecu}{\mathbf{u}}
\newcommand{\veca}{\mathbf{a}}
\newcommand{\vecb}{\mathbf{b}}
\newcommand{\vecv}{\mathbf{v}}
\newcommand{\vecz}{\mathbf{z}}

\numberwithin{equation}{section}

\title{%
  \texorpdfstring{$\SVP_\infty$}{SVP∞} Is NP-Hard for Power-of-two Cyclotomic Ideals%
}

\titlerunning{SVP in Power-of-two Cyclotomic Ideals}

\ifnum\fullversion=1\relax

  \author{Yansong Feng \and Jiaqi Liu \and Hengyi Luo \and Yanbin Pan}

  \authorrunning{Y. Feng, J. Liu, H. Luo, and Y. Pan}

  \institute{%
    Academy of Mathematics and Systems Science, Chinese Academy of Sciences\\
    \email{\{fengyansong,ljqi,luohengyi,panyanbin\}@amss.ac.cn}%
  }

\else

  \author{Anonymous Authors}
  \authorrunning{Anonymous Authors}
  \institute{}

\fi

\begin{document}

% All current revisions have been accepted;
% mark future edits with \change and \deleted.

\pagestyle{plain}
\maketitle
\thispagestyle{plain}

\begin{abstract}
We prove that exact decision-$\SVP$ in the coefficient
$\ell_\infty$-norm is $\NP$-complete for principal ideals and free
modules of rank two over the cyclotomic rings
$R_n=\Z[X]/(X^n+1)$, where $n$ ranges over powers of two. For each case, we give a deterministic
polynomial-time many-one reduction from the \SSum\ problem. For an instance with $s$ positive input integers, the
output lattice has dimension $O(s)$. A solution to the original
\SSum\ instance can be recovered in polynomial time from any
nonzero output lattice vector of norm at most the specified bound.

We also transfer these reductions from $R_n$ to the power-of-two
cyclic rings $S_{2n}=\Z[X]/(X^{2n}-1)$. This gives
$\NP$-completeness of decision-$\SVP$ in the coefficient
$\ell_\infty$-norm for principal ideals and free modules of any
fixed positive rank over these rings.
\end{abstract}

% ============================================================
% Main text
% ============================================================
% Section 1 review edits accepted; future edits remain marked in red.
\section{Introduction}\label{sec:introduction}
A lattice $\mathcal{L}\subseteq\R^m$ is the set of all integer linear combinations of linearly independent vectors in $\R^m$. Given a basis of $\mathcal{L}$ and a threshold $T>0$, the decision shortest vector problem ($\SVP$) asks whether $\mathcal{L}$ contains a nonzero vector of norm at most $T$. For general integer lattices, van Emde Boas proved deterministic $\NP$-hardness of $\SVP$ in the $\ell_\infty$-norm~\cite{vEB}. Dinur later proved $\NP$-hardness of approximating $\SVP$ in the $\ell_\infty$-norm within almost-polynomial factors~\cite{Dinur}. In the $\ell_2$-norm, Ajtai proved randomized $\NP$-hardness~\cite{Ajtai}, and Micciancio obtained randomized hardness of approximation within constant factors below $\sqrt2$~\cite{Micciancio}. Khot extended this randomized $\NP$-hardness to approximation within any fixed constant factor~\cite{Khot}. Wan later proved deterministic $\NP$-hardness of approximation within any constant factor greater than $1$~\cite{WanEuclideanApprox}.

The worst-case hardness of $\SVP$ also provides a basis for lattice-based cryptography. Worst-case-to-average-case reductions connect lattice approximation problems to the average-case learning with errors (LWE) problem. Regev~\cite{Regev} gave quantum reductions from worst-case $\GapSVP$ (the gap decision version of $\SVP$) and the shortest independent vectors problem ($\mathsf{SIVP}$) to LWE. Both source problems have polynomial approximation factors in the $\ell_2$-norm. Peikert~\cite{PeikertNorms} extended these quantum reductions to $\GapSVP$ and $\mathsf{SIVP}$ in the $\ell_\infty$-norm. Peikert~\cite{PeikertClassical} also gave a classical reduction from $\GapSVP$ in the $\ell_\infty$-norm to LWE for exponentially large moduli. Brakerski et al.~\cite{BLPRS} later gave a classical reduction from worst-case $\GapSVP$ in the $\ell_2$-norm to LWE with polynomially large moduli, at the cost of increasing the dimension.

Cryptographic schemes often use lattices with additional algebraic structure. Under the coefficient embedding, ideals and modules over cyclotomic rings give ideal and module lattices that are closed under multiplication by ring elements. The ring structure supports compact descriptions and fast polynomial arithmetic. Early constructions of collision-resistant hash functions were based on worst-case approximation assumptions for ideal lattices~\cite{LM} and cyclic lattices~\cite{PR}. Worst-case approximation problems on ideal and module lattices over cyclotomic rings also admit reductions to Ring-LWE~\cite{LPR} and Module-LWE~\cite{LS}, respectively. The security of ML-KEM, standardized by NIST in FIPS~203, is based on the hardness of Module-LWE over a power-of-two cyclotomic ring~\cite{FIPS203}.

Several recent works have established hardness of $\SVP$ in structured settings. Martin gives dimension-preserving reductions from general Euclidean lattice problems to the canonical embeddings of invertible ideals in totally real monogenic orders~\cite{Martin}. Wan proves deterministic $\NP$-hardness of coefficient Euclidean $\SVP$ on full-rank ideals of $\Z[X]/(X^N-1)$, where $N=q-1$ for an odd prime $q$~\cite{Wan}. Liu, Feng, and Pan prove deterministic $\NP$-completeness of canonical Euclidean $\SVP$ on  free modules of rank two over prime-cyclotomic rings with prime conductor $q\equiv3\pmod4$~\cite{LFP}.

We ask the following two questions.
\begin{quote}

\begin{enumerate}
\item \emph{Is $\SVP$ in the coefficient $\ell_\infty$-norm $\NP$-complete for principal ideals and free modules over power-of-two cyclotomic rings?}
\item \emph{Is $\SVP$ in the coefficient $\ell_\infty$-norm $\NP$-complete for principal ideals and free modules over power-of-two cyclic rings?}
\end{enumerate}
\end{quote}

\subsection{Our results}\label{sec:intro-results}
We give deterministic polynomial-time many-one reductions from the \SSum\ problem to exact $\SVP$ in the coefficient $\ell_\infty$-norm over the cyclotomic rings
\[
 R_n=\Z[X]/(X^n+1)
\]
and the cyclic rings $S_{2n}=\Z[X]/(X^{2n}-1)$, where $n$ ranges over powers of two.

\begin{theorem}[Informal]\label{thm:intro-cyclotomic}
$\SVP$ in the coefficient $\ell_\infty$-norm is $\NP$-complete for  free modules of rank two and for principal ideals over $R_n$, where $n$ ranges over powers of two.
\end{theorem}

The principal-ideal reduction also gives $\NP$-completeness for free modules of any fixed positive rank over $R_n$. The separate rank-two construction has smaller lattice dimension and gives a direct polynomial encoding of \SSum. Its two-coordinate subset-sum check motivates the principal-ideal construction.

We also extend the cyclotomic-to-cyclic construction of Liu, Feng, and Pan~\cite{LFPCVP} to free modules of arbitrary fixed rank, obtaining an exact relation between the first minima in the coefficient $\ell_\infty$-norm.
\begin{theorem}[Informal]\label{thm:intro-cyclic}
$\SVP$ in the coefficient $\ell_\infty$-norm is $\NP$-complete for principal ideals and free modules of any fixed positive rank over $S_{2n}$, where $n$ ranges over powers of two.
\end{theorem}

For a \SSum\ instance with $s$ positive input integers, the two cyclotomic reductions for principal ideals and free rank-two modules over $R_n$ each produce a lattice of dimension $O(s)$. In either reduction, a solution to the \SSum\ instance can be recovered in deterministic polynomial time from any nonzero vector in the constructed lattice whose $\ell_\infty$-norm is at most the specified bound. The corresponding search problems are therefore $\NP$-hard under polynomial-time Turing reductions. The cyclic transfer gives the same $O(s)$ dimension bound and search hardness for every fixed module rank over $S_{2n}$.

\subsection{Technical overview}\label{sec:intro-overview}
Our reductions start from the \SSum\ problem, whose input consists of positive integers $w_1,\ldots,w_s$ and a nonnegative integer target $B$. The problem asks whether there are Boolean choices $\xi_j\in\{0,1\}$ satisfying $\sum_{j=1}^s w_j\xi_j=B$. To encode this condition, each reduction constructs a lattice $\mathcal L$ and a positive integer $T$ such that
\[
 \exists\,{\vecxi}\in\{0,1\}^s:\ \sum_{j=1}^s w_j\xi_j=B
 \quad\Longleftrightarrow\quad
 \exists\,\mathbf0\ne\vecy\in \mathcal L:\ \norm{\vecy}_\infty\le T.
\]
We begin with an integer lattice, adapt its encoding to rank-two modules and principal ideals over $R_n$, and then transfer the constructions to cyclic rings. \par\medskip\noindent\textbf{\boldmath Warm-up: $\ell_\infty$-norm $\NP$-hardness in general lattices.} We first recall a reduction from the \SSum\ problem~\cite{Karp} to $\SVP$ in general lattices, illustrating the classical $\NP$-hardness result for the $\ell_\infty$-norm~\cite{vEB}. To encode the Boolean restriction, we represent $\xi_j$ by an integer variable $a_j$ and require $|2a_j-1|\le1$, which is equivalent to $a_j\in\{0,1\}$. To encode this condition in a lattice, we replace the fixed constant $1$ in $2a_j-1$ by an integer variable $b$. This replacement gives the linear coordinate $2a_j-b$, which agrees with $2a_j-1$ when $b=1$. We will force nonzero vectors of $\ell_\infty$-norm at most $1$ to have $b=\pm1$, and then take $b=1$ by negating the vector if necessary.

Put $W\coloneqq\sum_{j=1}^s w_j$. To test the subset-sum equation, we add the coordinate $\sum_{j=1}^s w_ja_j-Bb$. We multiply it by $2$ so that an $\ell_\infty$-norm bound of $1$ forces it to be zero. To obtain a full-rank lattice, we introduce an integer variable $c$ and add $Cc$ to this coordinate, where $C\coloneqq2(W+B+1)$. The even value of $C$ preserves this parity condition, and its size will force $c=0$ for short vectors. Together with the coordinates $2a_j-b$ and $b$, this gives the lattice
\begin{equation}\label{eq:overview-integer-lattice}
 \left\{
 \begin{pmatrix}
  2\veca-b\mathbf1\\
  b\\
  2\bigl(\sum_{j=1}^s w_ja_j-Bb\bigr)+Cc
 \end{pmatrix}
 : \veca\in\Z^s,\ b,c\in\Z
 \right\}\subseteq\Z^{s+2},
\end{equation}
where $\mathbf1\coloneqq(1,\ldots,1)^{\mathsf T}\in\Z^s$.

A solution $\vecxi$ to the \SSum\ instance gives a nonzero vector of $\ell_\infty$-norm $1$ in~\eqref{eq:overview-integer-lattice} by taking $\veca=\vecxi$, $b=1$, and $c=0$. Conversely, for a nonzero vector in~\eqref{eq:overview-integer-lattice} of $\ell_\infty$-norm at most $1$, the second coordinate gives $b\in\{0,\pm1\}$. The case $b=0$ would force $\veca=\mathbf0$ in the first block and $c=0$ in the last coordinate, giving the zero vector. Thus $b=\pm1$, and after negating the vector if necessary we may assume $b=1$. The first block then gives $a_j\in\{0,1\}$.

With these Boolean choices, the last coordinate tests the subset-sum equation. It is even and has absolute value at most $1$, so $2(\sum_{j=1}^s w_ja_j-B)+Cc=0$. The Boolean conditions give $0\le\sum_{j=1}^s w_ja_j\le W$. Since
\[
 |Cc|=\left|2\left(\sum_{j=1}^s w_ja_j-B\right)\right|
 \le2(W+B)<C,
\]
and $c\in\Z$, we obtain $c=0$ and $\sum_{j=1}^s w_ja_j=B$. Thus the lattice in~\eqref{eq:overview-integer-lattice} contains a nonzero vector of $\ell_\infty$-norm at most $1$ if and only if the \SSum\ instance has a solution.

\par\medskip\noindent\textbf{Encoding Subset Sum in rank-two modules.} The lattice in~\eqref{eq:overview-integer-lattice} encodes the subset-sum equation using integer variables. Similarly, we construct a rank-two module over the cyclotomic ring $R_n\coloneqq\Z[X]/(X^n+1)$, where $n$ is a power of two. We store the Boolean choices in the coefficients of a polynomial $a\in R_n$. The integer variable $b$ is likewise replaced by a polynomial $b\in R_n$. However, $\norm b_\infty\le1$ allows many nonzero terms (for example, $b=1+X+\cdots+X^{n-1}$), so this bound alone does not give the normalization to $b=1$ used in the integer construction. We therefore need an additional constraint forcing a nonzero $b$ to be a signed monomial $b=\pm X^\ell$. Multiplication by $b^{-1}$ then preserves the coefficient $\ell_\infty$-norm and reduces to $b=1$.

\par\smallskip\noindent\emph{Step 1. Encoding Boolean choices.} We assume $0\le B\le W$. Let $n\coloneqq2m$ with $m>s$ a power of two and put $\polyu\coloneqq\sum_{j=1}^sX^{j-1}$. Here $\polyu$ plays the role of the all-ones vector $\mathbf1$, so $z\coloneqq2a-\polyu b$ is the polynomial counterpart of the first lattice block $2\veca-b\mathbf1$ in~\eqref{eq:overview-integer-lattice}. We first choose the second module coordinate to be $Mz$ and the $\ell_\infty$-norm threshold to be $M$, where $M$ is a positive integer to be fixed later. Every vector of $\ell_\infty$-norm at most $M$ then satisfies $\norm z_\infty\le1$. If $b=0$, then $z=2a$, and the second-coordinate bound gives $\norm{2a}_\infty\le1$, forcing $a=0$ by integrality. Thus every nonzero vector in this norm range has $b\ne0$, as in the integer construction~\eqref{eq:overview-integer-lattice}.

We first describe the Boolean and subset-sum checks for the normalized case $b=1$. In this case, the second-coordinate bound becomes $\norm{2a-\polyu}_\infty\le1$ and forces $a$ to have Boolean coefficients, with $a=\sum_{j=1}^s\xi_jX^{j-1}$ and $\xi_j\in\{0,1\}$.

\par\smallskip\noindent\emph{Step 2. Checking the subset-sum equation.} For $b=1$, we encode the weights in the \SSum\ instance by setting $\polyhbase\coloneqq\sum_{j=1}^s w_jX^{-(j-1)}$. The opposite exponents make the constant coefficient of $\polyhbase z$, denoted by $[\polyhbase z]_0$, equal to $2\sum_{j=1}^s w_j\xi_j-W$. To compare $\sum_{j=1}^s w_j\xi_j$ with the target $B$, we need to compare this constant coefficient with $2B-W$.

We choose the first coordinate of the module lattice so that a nonzero difference $[\polyhbase z]_0-(2B-W)$ makes at least one of its coefficients exceed $M$ in absolute value. To achieve this goal, we give the constant and $X^m$ coefficients the values $-M$ plus and minus this difference, respectively. Since $s<m$, the coefficient of $X^m$ in $\polyhbase z$ is zero. Multiplying $\polyhbase z-(2B-W)$ by $1-X^m$ then keeps this difference in the constant coefficient and shifts its negative to the coefficient of $X^m$. We therefore set
\[
 \polyh\coloneqq(1-X^m)\polyhbase,\qquad
 \polyv\coloneqq(2B-W)(1-X^m).
\]
Subtracting $M(1+X^m)$ from $\polyh z-\polyv$ gives the constant and $X^m$ coefficients
\[
 -M+\bigl([\polyhbase z]_0-(2B-W)\bigr),\qquad
 -M-\bigl([\polyhbase z]_0-(2B-W)\bigr).
\]
The two coefficients are symmetric about $-M$. If $[\polyhbase z]_0\ne2B-W$, either the constant coefficient or the coefficient of $X^m$ is less than $-M$ and therefore has absolute value greater than $M$. Thus a vector of $\ell_\infty$-norm at most $M$ must satisfy the subset-sum equation.

\par\smallskip\noindent\emph{Step 3. Restricting $b$ to a signed monomial.} We now return to general $b$ and first take $\polyh z-(M\polyd+\polyv)b$ as the first coordinate, where $\polyd\coloneqq1+X^m$. A vector of $\ell_\infty$-norm at most $M$ then satisfies $\norm{\polyh z-(M\polyd+\polyv)b}_\infty\le M$. Multiplication by $\polyd$ replaces the coefficients of $b$ at indices $j,j+m$ by their difference and sum, whose maximum absolute value equals the sum of the original absolute values. Together with $\norm z_\infty\le1$, this first-coordinate bound forces $\norm{\polyd b}_\infty=1$ for $b\ne0$ when $M$ is sufficiently large. Consequently, for each $0\le j<m$, at most one of the coefficients of $X^j$ and $X^{j+m}$ in $b$ is nonzero, and any nonzero coefficient equals $1$ or $-1$. Every nonzero coefficient of $M\polyd b$ therefore has absolute value $M$. However, the condition $\norm{\polyd b}_\infty=1$ still allows $b$ to have several nonzero coefficients (for example, $b=1+X$), so an additional constraint is needed to force $b$ to be a signed monomial.

To exclude these remaining possibilities, we set $\polyq\coloneqq M\polyd+L\polyd\polyk$, with $\polyk\in R_n$ and a positive integer $L$. The completed construction gives a free module of rank two whose vectors have the following form with $a,b\in R_n$
\[
 \binom{\polyh z-(\polyq+\polyv)b}{Mz}.
\]
For sufficiently large $M$, every nonzero vector of this module with $\ell_\infty$-norm at most $M$ satisfies $\norm{\polyd b}_\infty=1$. The polynomial $\polyk$ is designed so that, if $\norm{\polyd b}_\infty=1$ and $b$ has more than one nonzero term, adding $L\polyd\polyk b$ to $M\polyd b$ increases the absolute value of at least one coefficient from $M$ to at least $M+L$. The remaining term satisfies $\norm{\polyh z-\polyv b}_\infty\le4W$. Choosing $L>4W$ therefore makes the first coordinate have $\ell_\infty$-norm greater than $M$ (Lemma~\ref{lem:module-monomial}).

Thus every nonzero vector of $\ell_\infty$-norm at most $M$ has $b=\pm X^\ell$. Multiplication by $b^{-1}$ preserves this norm and reduces to $b=1$. The constant and $X^m$ coefficients of $L\polyd\polyk$ are zero, so the constant coefficient and the coefficient of $X^m$ in $\polyq$ both equal $M$. The second-coordinate bound then gives the Boolean coefficients of $a$, and the first-coordinate bound forces $\sum_{j=1}^s w_j\xi_j=B$. Conversely, a solution to the \SSum\ instance gives a vector of $\ell_\infty$-norm $M$, since the constant and $X^m$ coefficients of the first coordinate are $-M$, its other coefficients have absolute value less than $M$, and $\norm{Mz}_\infty=M$.

\par\medskip\noindent \textbf{Encoding Subset Sum in principal ideals.} In the rank-two construction, $a$ stores the Boolean choices, while a signed monomial $b$ is normalized to $1$ for the subset-sum check. A vector in a principal ideal has the form $f\sigma$, so the same polynomial $\sigma\in R_n$ must encode the Boolean choices and at the same time allow us to check the subset-sum equation.

\par\smallskip\noindent\emph{Step 1. Encoding Boolean choices.} Let $m$ be a sufficiently large power of two and set $n\coloneqq2m$. We encode each Boolean choice $\xi_j$ by the position of a nonzero coefficient of $\sigma$ in the three-element set $\{8j-1,8j,8j+1\}$. The center $8j$ represents $\xi_j=0$, and either $8j-1$ or $8j+1$ represents $\xi_j=1$. To construct a polynomial from given Boolean choices $\vecxi\in\{0,1\}^s$, we take the index $8j+\xi_j$ and add the constant term $1$, giving
\[
\sigma\coloneqq1+\sum_{j=1}^{s}X^{8j+\xi_{j}}.
\]
The constant term fixes $\sigma_0=1$, allowing the target $B$ to enter the subset-sum check as a fixed subtraction.

\par\smallskip\noindent\emph{Step 2. Checking the subset-sum equation.} We retain the module construction's subset-sum check, in which the difference enters two ordinary coefficients with opposite signs. In the ring setting, write $\ii=X^m$ and group the ordinary coefficients at $\ell$ and $\ell+m$ into the complex coefficient $\sigma_\ell=[\sigma]_\ell+\ii[\sigma]_{\ell+m}$ for $0\le\ell<m$. Write $Z\coloneqq\sum_{j=1}^{s}w_j\xi_j-B$ for the subset-sum error associated with the Boolean choices $\vecxi$. Let $T>0$ be the $\ell_\infty$-norm bound. We first seek a polynomial $\polyres\in R_n$ for which $(\polyres\sigma)_0=\ii Z$. Therefore, for the generator $(T-\polyres)\polyd$ with $\polyd=1+\ii$, we have
\[
\bigl((T-\polyres)\polyd\sigma\bigr)_0=(1+\ii)(T-\ii Z)=(T+Z)+\ii(T-Z).
\]
The larger absolute value of these two ordinary coefficients is $T+|Z|$, so the $\ell_\infty$-norm bound $T$ enforces $Z=0$.

Just as $\polyhbase$ reads the coefficients of $z$ through the constant coefficient of $\polyhbase z$, we construct a polynomial $\polyres$ to read the choices encoded in $\sigma$ through the complex coefficient $(\polyres\sigma)_0$. Multiplication by $X^{-r}$ makes $\sigma_r$ contribute to the complex coefficient at $0$. To guarantee that each set $\{8j-1,8j,8j+1\}$ with $1\le j\le s$ contains a nonzero complex coefficient of $\sigma$ under the norm bound, we introduce a common offset $H\coloneqq W+1$. We assign the value $H$ to the center $8j$ and $H+w_j$ to the two adjacent indices $8j-1$ and $8j+1$. Since $H>W$, omitting an entire set loses more than all the input weights can supply.

For the polynomial $\sigma$ constructed in Step~1, subtracting $sH+B$ from $sH+\sum_{j=1}^s w_j\xi_j$ gives $Z$. We want the ordinary coefficients at indices $0$ and $m$ to form the pair $(T+Z,T-Z)$. To obtain these coefficients after multiplication by $\polyd$, we place $Z$ in the imaginary part of $(\polyres\sigma)_0$ by defining
\[
\polyres\coloneqq\ii\left(
\sum_{j=1}^{s}\left[
HX^{-8j}+(H+w_{j})\bigl(X^{-(8j-1)}+X^{-(8j+1)}\bigr)
\right]-(sH+B)
\right).
\]
Thus $(\polyres\sigma)_0=\ii Z$.

\par\smallskip\noindent\emph{Step 3. Correcting the generator.} However, the vector $(T-\polyres)\polyd\sigma$ violates the norm bound in the ordinary coefficient pair at indices $p_s$ and $p_s+m$, where $p_s\coloneqq8s+\xi_s$ is the index of the last nonzero complex coefficient of $\sigma$. The terms containing negative powers of $X$ in the definition of $\polyres$ contribute zero at $p_s$, so $(\polyres\sigma)_{p_s}=-\ii(sH+B)$. The ordinary coefficients at $p_s$ and $p_s+m$ consequently have maximum absolute value $T+sH+B>T$, even when $Z=0$. For $\sigma$ constructed in Step~1, we design the correction $(\polyg-\polyres)\sigma$ to have real coefficient $C$ at index $0$ and real coefficients at least $2C$ at the other indices where $\sigma$ has a nonzero complex coefficient, including $p_s$. We choose $C\coloneqq\norm{\polyres}_1+1$ and raise the main term to $(T+C)\polyd\sigma$. The correction is designed to bring both ordinary coefficients of $f\sigma$ below $T$ in absolute value at every other nonzero complex position of $\sigma$. Thus $(\polyg\sigma)_0=C+\ii Z$, and $(T+C-\polyg)\sigma$ has complex coefficient $T-\ii Z$ at $0$. We therefore take the generator $f\coloneqq(T+C-\polyg)\polyd$. Multiplication by $\polyd$ gives
\[
\polyd(T-\ii Z)=(T+Z)+\ii(T-Z).
\]

\par\smallskip\noindent\emph{Step 4. Enforcing the encoding.} It remains to control the positions and complex coefficients of $\sigma$ for every nonzero vector $f\sigma$ of $\ell_\infty$-norm at most $T$. To enforce the intended encoding, we take $\polyg$ to be of the form
\begin{equation}\label{eq:overview-principal-g}
\polyg\coloneqq10C\polyae+\polytzero+\polyres.
\end{equation}
With $T\coloneqq3\norm{\polyg}_1+1$, we can prove that every nonzero vector $f\sigma$ of $\ell_\infty$-norm at most $T$ satisfies $\norm{\polyd\sigma}_\infty=1$. Hence the complex coefficients of $\sigma$ lie in $\{0,\pm1,\pm\ii\}$. We explain the roles of the three terms under this norm bound.

Multiplication by a power of $X$ preserves $\norm{f\sigma}_\infty$. We therefore use the dominant term $10C\polyae$ to constrain the pairwise cyclic distances between the indices of nonzero complex coefficients instead of their absolute positions. We build $\polyae$ from differences $X^\delta-X^{-\delta}$ for distances outside $E\coloneqq\bigcup_{j=1}^s\{8j-1,8j,8j+1\}$. The norm bound forces all pairwise cyclic distances to belong to $E$ and $(\polyae\sigma)_\ell=0$ whenever $\sigma_\ell\ne0$ (Lemma~\ref{lem:gaussian-distances}). After multiplying by a suitable power of $X$, we may assume $\sigma_0=1$. Every other nonzero complex coefficient then has its index in one of the three-element sets, with at most one such index in each set.

The term $\polyres$, defined above, uses the common offset $H>W$ to ensure that each set $\{8j-1,8j,8j+1\}$ contains a nonzero complex coefficient of $\sigma$. A missing copy of $H$ cannot be compensated by the remaining input weights. Together with the distance restrictions from $\polyae$, the norm bound gives exactly one nonzero complex coefficient in each set $\{8j-1,8j,8j+1\}$. For the polynomial constructed in Step~1, $\polytzero$ supplies the correction described in Step~3. With $\polytzero$, the norm bound then forces all nonzero complex coefficients of $\sigma$ to equal $1$ (Lemma~\ref{lem:principal-support}).

Reading the Boolean choices from the nonzero positions of $\sigma$ as in Step~1, we obtain $(\polyres\sigma)_0=\ii Z$. Combining this with $(\polyae\sigma)_0=0$ and $(\polytzero\sigma)_0=C$ gives $(\polyg\sigma)_0=C+\ii Z$. The ordinary coefficients of $f\sigma$ at $0$ and $m$ are therefore $T+Z$ and $T-Z$, so the norm bound forces $Z=0$. Conversely, a solution gives a polynomial $\sigma$ of the displayed form for which every ordinary coefficient of $f\sigma$ has absolute value at most $T$, with equality at indices $0$ and $m$. The first minimum is therefore $T$ on YES instances and at least $T+1$ on NO instances by integrality.

\par\medskip\noindent\textbf{Transfer to cyclic rings.} The preceding reductions use $R_n=\Z[X]/(X^n+1)$, where $n$ is a power of two. Multiplication by $X$ gives a negacyclic shift in $R_n$ and a cyclic shift in $S_{2n}=\Z[X]/(X^{2n}-1)$. The $\Z$-linear embedding $p\mapsto(1-X^n)p$ sends the coefficient vector $\mathbf p\in\Z^n$ of $p$ to $(\mathbf p,-\mathbf p)$. The identity $(1-X^n)(X^n+1)=0$ in $S_{2n}$ ensures that multiplication by $X$ in $S_{2n}$ gives the required negacyclic shift on the first block. We therefore seek  a principal ideal of $S_{2n}$ whose shortest vectors have the form $(\mathbf w,-\mathbf w)$ with $\mathbf w\in\Z^n$.

We take $r=1$ for example. We follow the construction of Liu, Feng, and Pan~\cite{LFPCVP}. Given $0\ne f\in R_n$, represented by a polynomial of degree less than $n$, take $C\coloneqq2\norm f_\infty+1$ and define
\[
\widehat f\coloneqq(1-X^n)f+C(1+X^n)\in S_{2n}.
\]

To analyze the coefficients of $\widehat f u$, write $u=a+X^n b\in S_{2n}$ with $a,b\in\Z[X]$ of degree less than $n$. Let $y$ be the representative of $f(a-b)\in R_n$ of degree less than $n$. Then
\[
\widehat f u=\bigl(y+C(a+b)\bigr)+X^n\bigl(-y+C(a+b)\bigr).
\]
The same polynomial $C(a+b)$ occurs in both blocks, and $y$ occurs with opposite signs. A nonzero coefficient of $C(a+b)$ cannot be canceled in both blocks, so $a+b\ne0$ forces $\norm{\widehat f u}_\infty\ge C$. On the other hand, $\widehat f(1-X^n)=2(1-X^n)f$ has norm $2\norm f_\infty<C$. Every shortest output vector must therefore have $b=-a$. With $b=-a$, the two output blocks are $2fa$ and $-2fa$, where $fa$ is reduced modulo $X^n+1$. Every vector $fa\in(f)$ has such an image by taking $u=a-X^n a$. Thus $\lambda_{1,\infty}((\widehat f))=2\lambda_{1,\infty}((f))$.

Section~\ref{sec:cyclic}  proves that the output module is free and extends this construction to polynomial matrices. It gives a deterministic polynomial-time reduction from $\SVP$ on free rank-$r$ modules over $R_n$ to the same problem over $S_{2n}$, for every fixed $r\ge1$.

\subsection{Organization}
Section~\ref{sec:prelim} establishes the algebraic preliminaries, coefficient norms, input representations, and complex coefficient representation. Section~\ref{sec:module} presents the rank-two module reduction. Section~\ref{sec:principal} presents the principal-ideal reduction. Section~\ref{sec:cyclic} transfers both reductions to cyclic polynomial rings, and Section~\ref{sec:conclusion} concludes. A Lean formalization of our results is available at \url{https://github.com/fffmath/pow2-cyclotomic-svp}.

\section{Preliminaries}\label{sec:prelim}
\noindent\textbf{Notation.} We write $\Z$, $\Q$, $\R$, and $\C$ for the integers, rationals, reals, and complex numbers, respectively. Polynomials are denoted by lowercase italic letters, vectors by bold lowercase letters, and matrices by bold uppercase letters. The superscript $\mathsf T$ denotes transpose. For $z\in\C$, $\overline z$, $\Re z$, and $\Im z$ denote its complex conjugate, real part, and imaginary part, respectively. Logarithms are to base two.

\subsection{Lattices}

Let $1\le d\le N$ and let $\matB=(\mathbf b_1,\ldots,\mathbf b_d)\in\R^{N\times d}$ have linearly independent columns. The \emph{lattice} generated by $\matB$ is
\[
 \mathcal L\coloneqq\matB\Z^d
 =\left\{\sum_{j=1}^d u_j\mathbf b_j:u_1,\ldots,u_d\in\Z\right\}.
\]
We call the columns of $\matB$ a \emph{basis} of $\mathcal L$ and $d$ the \emph{rank} of $\mathcal L$~\cite{Micciancio}. We call $\mathcal L$ \emph{full-rank} if $d=N$.

For $\mathbf x=(x_1,\ldots,x_N)^{\mathsf T}\in\R^N$, define its \emph{$\ell_\infty$-norm} by $\norm{\mathbf x}_\infty\coloneqq\max_{1\le j\le N}|x_j|$. We define the \emph{first minimum} of $\mathcal L$ in this norm by
\[
 \lambda_{1,\infty}(\mathcal L)\coloneqq
 \min_{\mathbf0\ne\mathbf x\in \mathcal L}\norm{\mathbf x}_\infty.
\]
Given an integer basis of a lattice $\mathcal L$ and a positive integer $\radius$, the \emph{decision shortest vector problem} in the $\ell_\infty$-norm asks whether $\lambda_{1,\infty}(\mathcal L)\le\radius$. Given an integer basis of $\mathcal L$, \emph{search-$\SVP$} in the $\ell_\infty$-norm asks for $\mathbf0\ne\mathbf x\in\mathcal L$ with $\norm{\mathbf x}_\infty=\lambda_{1,\infty}(\mathcal L)$.

\subsection{Cyclotomic ring}

Let $n\ge1$ be a power of two and set
\[
 R_n\coloneqq\Z[X]/(X^n+1).
\]
Let $\zeta\coloneqq e^{\pi\ii/n}$ be a primitive $2n$-th root of unity. Since $X^n+1$ is the $2n$-th cyclotomic polynomial, evaluation at $\zeta$ induces the isomorphism $R_n\cong\Z[\zeta]$ sending $X$ to $\zeta$. The ring $\Z[\zeta]$ is the full ring of integers of the cyclotomic field $\Q(\zeta)$~\cite{Washington}. Hence $R_n$ is an integral domain. We identify each element with its unique representative of degree less than $n$. For $g\in R_n$ and $0\le\ell<n$, $[g]_\ell\in\Z$ denotes the \emph{ordinary coefficient} of $X^\ell$ in this representative. The \emph{coefficient embedding} with respect to the power basis $1,X,\ldots,X^{n-1}$ maps $g=\sum_{\ell=0}^{n-1}[g]_\ell X^\ell\in R_n$ to $([g]_0,\ldots,[g]_{n-1})^{\mathsf T}\in\Z^n$. We identify $g$ with this vector and represent tuples of polynomials by concatenating their coefficient vectors.

For $g\in R_n$, define its norms by its coefficient vector:
\[
 \norm g_1\coloneqq\sum_{\ell=0}^{n-1}|[g]_\ell|,
 \qquad
 \norm g_\infty\coloneqq\max_{0\le\ell<n}|[g]_\ell|.
\]
For $\vecz=(z_1,\ldots,z_r)\in R_n^r$, set $\norm{\vecz}_\infty\coloneqq\max_{1\le a\le r}\norm{z_a}_\infty$. For $g,z\in R_n$, we have
\begin{equation}\label{eq:product-bound}
 \norm{gz}_\infty\le\norm g_1\norm z_\infty.
\end{equation}

Multiplication by $g\in R_n$ is represented in the power basis by the matrix
\begin{equation}\label{eq:action}
 \matM_g\coloneqq\bigl([X^jg]_\ell\bigr)_{0\le\ell,j<n}.
\end{equation}
Its $j$-th column is the coefficient vector of $X^jg$ for $0\le j<n$. Since $X^n=-1$, multiplication by $X$ permutes coefficients up to sign, and $X^{-1}=-X^{n-1}$.

For $g,z\in R_n$, their \emph{coefficient inner product} is $\ip gz\coloneqq\sum_{\ell=0}^{n-1}[g]_\ell[z]_\ell$. For $g\in R_n$, define its \emph{adjoint} by $g^*\coloneqq g(X^{-1})$. Under the identification $R_n\cong\Z[\zeta]$, this operation corresponds to complex conjugation. For $g,z,w\in R_n$, we have
\begin{equation}\label{eq:adjoint-bound}
 \ip{gz}w=\ip z{g^*w}.
\end{equation}
The inner-product identity follows by linearity because multiplication by $X^{-1}$ is the transpose of multiplication by $X$. A polynomial $g$ is \emph{skew-adjoint} if $g^*=-g$. In this case, $\ip z{gz}=\ip z{g^*z}=-\ip z{gz}$, so $\ip z{gz}=0$.

\subsection{Ideal and module lattices}\label{subsec:ideal-module-svp}
An \emph{ideal} $\mathcal I\subseteq R_n$ is an additive subgroup closed under multiplication by elements of $R_n$. More generally, for $r\ge1$, an $R_n$-\emph{submodule} $\mathcal M\subseteq R_n^r$ is an additive subgroup closed under scalar multiplication by elements of $R_n$. Under the coefficient embedding, a nonzero ideal of $R_n$ gives an integer lattice in $\Z^n$, called an \emph{ideal lattice}~\cite{LPR}. For $0\ne f\in R_n$, the \emph{principal ideal} $(f)\coloneqq\{fa:a\in R_n\}$ has coefficient lattice $\matM_f\Z^n$.  A nonzero $R_n$-submodule of $R_n^r$ similarly gives a \emph{module lattice} in $\Z^{rn}$. We call an $R_n$-module \emph{free of rank $r$} if it has an $R_n$-basis with $r$ elements. We consider free modules  $\mathcal M\coloneqq\matG R_n^r$, where $\matG=(\polyg_{ab})\in R_n^{r\times r}$ and $\det\matG\ne0$. Their coefficient lattices have integer basis $\matB\coloneqq(\matM_{\polyg_{ab}})_{1\le a,b\le r}\in\Z^{rn\times rn}$.

We consider the following decision versions of $\SVP$ in the coefficient $\ell_\infty$-norm.
\begin{definition}[Ideal and module $\SVP$]\label{def:modules}
Given a power of two $n$, a nonzero ideal  $\mathcal I\subseteq R_n$, and a positive integer $\radius$, the problem $\ISVP_\infty$ asks whether  $\lambda_{1,\infty}(\mathcal I)\le\radius$.

Given a power of two $n$, a free rank-two $R_n$-module  $\mathcal M\subseteq R_n^2$, and a positive integer $\radius$, the problem $\MSVP_{\infty,2}$ asks whether  $\lambda_{1,\infty}(\mathcal M)\le\radius$.
\end{definition}

The corresponding \emph{search-$\ISVP_\infty$} and \emph{search-$\MSVP_{\infty,2}$} problems take $(n,\mathcal I)$ and $(n,\mathcal M)$, respectively, and ask for a nonzero vector of minimum coefficient $\ell_\infty$-norm in the given ideal or module.

In the problems above, the ideal $\mathcal I$ is specified by a $\Z$-basis in $R_n$, and the free module $\mathcal M$ by an $R_n$-basis matrix $\matG$. The coefficient embedding represents these basis elements as integer coefficient vectors. We use their ring representations below.

\begin{lemma}\label{lem:integer-operations}
The problems $\ISVP_\infty$ and $\MSVP_{\infty,2}$ belong to $\NP$.
\end{lemma}
\begin{proof}
For either problem, let $r\coloneqq1$ for ideals and $r\coloneqq2$ for modules. A YES certificate is a vector $\vecy\in\Z^{rn}$ with $1\le\norm{\vecy}_\infty\le\radius$. Its bit length is $O(rn\log(\radius+1))$. The verifier checks the input conditions and norm bound. It tests lattice membership by solving $\matB\vecu=\vecy$ over $\Q$ and checking $\vecu\in\Z^{rn}$,  using the coefficient vectors of the supplied ideal basis and the block expansion of $\matG$ for modules. For an ideal input, it also applies this test to $X$ times each basis vector, since $R_n$ is generated by $X$ over $\Z$. All these checks take polynomial time~\cite{Cohen}.\qed
\end{proof}

\subsection{The subset sum problem}
\begin{definition}[\SSum]\label{def:subset-sum}
An instance of \SSum\ consists of $s\ge1$ positive integer \emph{weights} $w_1,\ldots,w_s$ and a nonnegative \emph{target} integer $B$. It is a \emph{YES instance} if there exists a Boolean vector $\vecxi=(\xi_1,\ldots,\xi_s)\in\{0,1\}^s$ such that
\[
 \sum_{j=1}^s w_j\xi_j=B.
\]
\end{definition}
We write $W\coloneqq\sum_{j=1}^s w_j$ for the \emph{total weight} and set $\beta\coloneqq\lceil\log_2(2+\max\{B,w_1,\ldots,w_s\})\rceil$. \SSum\ is $\NP$-complete~\cite{Karp}.

\subsection{Complex coefficient representation}
Let $n=2m$, where $m\ge1$ is a power of two. Under the isomorphism $R_n\cong\Z[\zeta]$, the element $X^m$ corresponds to $\zeta^m=\ii$. We write $\ii=X^m$ in $R_n$. Since $\ii^2=-1$, every $z\in R_n$ has a unique representation
\begin{equation}\label{eq:gaussian-blocks}
 z=\sum_{j=0}^{m-1}z_jX^j,\qquad
 z_j\coloneqq[z]_j+\ii[z]_{j+m}\in\Z[\ii].
\end{equation}
We call $z_j$ the $j$-th \emph{complex coefficient} of $z$. The ordinary coefficient norm satisfies $\norm z_\infty=\max_{0\le j<m}\{ |\Re z_j|,|\Im z_j|\}$. For $z,w\in R_n$, the coefficient inner product satisfies $\ip zw=\Re\sum_{j=0}^{m-1}\overline{z_j}w_j$. The \emph{support} of the coefficient sequence $(z_0,\ldots,z_{m-1})$ is $\{j\in\{0,\ldots,m-1\}:z_j\ne0\}$. The units of $\Z[\ii]$ are $1,-1,\ii,-\ii$.

We also extend the complex coefficients to all integer indices by setting
\begin{equation}\label{eq:gaussian-extension}
 z_{j+qm}\coloneqq\ii^{-q}z_j
       \qquad(0\le j<m,\ q\in\Z).
\end{equation}
In particular, $z_{j+m}=-\ii z_j$, and $(X^az)_j=z_{j-a}$ for every integer $a$ and $0\le j<m$.

Set $\polyd\coloneqq1+\ii=1+X^m$. To describe the nonzero polynomials satisfying $\norm{\polyd b}_\infty=1$, define
\[
 \GU\coloneqq\left\{0\ne b\in R_n:
 b_j\in\{0,\pm1,\pm\ii\}\text{ for }0\le j<m\right\}.
\]
Thus $\GU$ consists of nonzero polynomials whose complex coefficients are zero or units of $\Z[\ii]$. Multiplication by $\polyd$ gives the following norm formula.

\begin{lemma}\label{lem:gaussian-unit}
For every $b\in R_n$, we have
\begin{equation}\label{eq:D-norms}
 \norm{\polyd b}_\infty
 =\max_{0\le j<m}\bigl(|\Re b_j|+|\Im b_j|\bigr).
\end{equation}
In particular, a nonzero $b\in R_n$ satisfies $\norm{\polyd b}_\infty=1$ if and only if $b\in\GU$. Every other nonzero $b$ satisfies $\norm{\polyd b}_\infty\ge2$.
\end{lemma}
\begin{proof}
For $0\le j<m$, multiplication by $\polyd=1+X^m$ gives
\[
 ([\polyd b]_j,[\polyd b]_{j+m})
 =([b]_j-[b]_{j+m},[b]_j+[b]_{j+m}).
\]
The identity $\max\{|c-e|,|c+e|\}=|c|+|e|$ for real $c,e$ therefore gives
\[
 \max\{|[\polyd b]_j|,|[\polyd b]_{j+m}|\}
 =|[b]_j|+|[b]_{j+m}|=|\Re b_j|+|\Im b_j|.
\]
Taking the maximum over $0\le j<m$ proves~\eqref{eq:D-norms}. Since $b_j\in\Z[\ii]$, the quantity $|\Re b_j|+|\Im b_j|$ is at most one exactly when $b_j\in\{0,\pm1,\pm\ii\}$. For nonzero $b$, the maximum is positive, so the remaining claims follow.\qed
\end{proof}

% Confirmed Section 3: existing notation is unmarked; mark future edits with \change{...}.
\section{Free modules of rank two}\label{sec:module}
In this section, we encode a \SSum\ instance $(w_1,\ldots,w_s;B)$ in a free rank-two module over $R_n$ whose nonzero vectors of $\ell_\infty$-norm at most $M$ determine solutions to the instance. The module vectors are designed to have the form
\begin{equation}\label{eq:module-vector-preview}
 \binom{\polyh z-(\polyq+\polyv)b}{Mz},
\end{equation}
where $z\coloneqq2a-\polyu b$ and $a,b\in R_n$. The polynomials $\polyu,\polyh,\polyv,\polyq$ and the positive integer $M$ will be specified in Section~\ref{subsec:module-construction}.

A vector of $\ell_\infty$-norm at most $M$ in the constructed lattice therefore satisfies $\norm z_\infty\le1$ and $\norm{\polyh z-(\polyq+\polyv)b}_\infty\le M$. We choose these polynomials so that every nonzero such vector has $b=\pm X^\ell$. Multiplication by $b^{-1}$ then preserves the coefficient $\ell_\infty$-norm and reduces to $b=1$. With this normalization, the second module coordinate forces $a$ to have Boolean coefficients, and the bounds on two coefficients of the first coordinate force $\sum_{j=1}^s w_j[a]_{j-1}=B$.

\subsection{A criterion for signed monomials}
Let $m\ge2$ be a power of two and set $n\coloneqq2m$. We recall that $\polyd=1+X^m$ and that the complex coefficients of $b\in R_n$ are $b_j=[b]_j+\ii[b]_{j+m}$, as in~\eqref{eq:gaussian-blocks}. When the vector in~\eqref{eq:module-vector-preview} has $\ell_\infty$-norm at most $M$, we first aim to force a nonzero $b$ to have a single nonzero complex coefficient.

For this purpose, define the polynomials
\begin{equation}\label{eq:two-moments}
 \begin{aligned}
 \polyazero&\coloneqq\sum_{\delta=1}^{m-1}(X^\delta-X^{-\delta})
             =(1+\ii)\sum_{\delta=1}^{m-1}X^\delta,\\
 {\polyaone}&\coloneqq\sum_{\delta=1}^{m-1}\delta(X^\delta-X^{-\delta})
             =\sum_{\delta=1}^{m-1}\bigl(\delta+\ii(m-\delta)\bigr)X^\delta.
 \end{aligned}
\end{equation}
where we used $X^{-\delta}=-\ii X^{m-\delta}$. The proof of the next lemma compares the first and last nonzero complex coefficients of $b$ using $\polyazero b$ and $\polyaone b$.

\begin{lemma}\label{lem:two-moments}
Let $0\ne b\in R_n$. If $(\polyazero b)_j=(\polyaone b)_j=0$ for every $0\le j<m$ with $b_j\ne0$, then $b$ has exactly one nonzero complex coefficient.
\end{lemma}
\begin{proof}
Suppose that $b$ has at least two nonzero complex coefficients, and let $j<r$ be the smallest and largest indices of its nonzero complex coefficients. By~\eqref{eq:two-moments} and~\eqref{eq:gaussian-extension},
\begin{equation}\label{eq:moment-zero}
 \begin{aligned}
  0&=(\polyazero b)_j=\ii(1+\ii)\sum_{\ell=j+1}^{r}b_\ell,\\
  0&=(\polyazero b)_r=(1+\ii)\sum_{\ell=j}^{r-1}b_\ell.
 \end{aligned}
\end{equation}
Both sums equal zero, and subtracting them gives $b_j=b_r$.

For $j<\ell\le r$, the $X^{m+j-\ell}$ term of $\polyaone$, together with $X^{m+j}=\ii X^j$, gives the coefficient $-(1+\ii)(\ell-j)+\ii m$ of $b_\ell$ in $(\polyaone b)_j$. For $j\le\ell<r$, the $X^{r-\ell}$ term gives the coefficient $(1-\ii)(r-\ell)+\ii m$ of $b_\ell$ in $(\polyaone b)_r$. The parts of the coefficients independent of $\ell$ multiply the zero sums in~\eqref{eq:moment-zero}, giving
\begin{equation}\label{eq:moment-one}
 \begin{aligned}
  0&=(\polyaone b)_j=-(1+\ii)\sum_{\ell=j+1}^{r}\ell b_\ell,\\
  0&=(\polyaone b)_r=-(1-\ii)\sum_{\ell=j}^{r-1}\ell b_\ell.
 \end{aligned}
\end{equation}
Subtracting the sums in~\eqref{eq:moment-one} gives $jb_j=rb_r$. Together with $b_j=b_r$, this yields $(r-j)b_j=0$, contradicting $r>j$ and $b_j\ne0$.\qed
\end{proof}

For $b\in R_n$ with $\norm b_\infty\le1$, the ordinary coefficients of $\polyazero b$ are uniformly bounded in terms of $m$. This allows us to choose an integer $\packbase$ so that the two zero-coefficient conditions in Lemma~\ref{lem:two-moments} can be expressed using a single polynomial. Define $\polyk\in R_n$ by
\begin{equation}\label{eq:packed-moments}
 \polyk\coloneqq\polyazero+\packbase\polyaone
 =\sum_{\delta=1}^{m-1}(1+\packbase\delta)(X^\delta-X^{-\delta}),
\end{equation}
where $\packbase\coloneqq2m+1$. This choice of $\packbase$ gives the following equivalence.

\begin{lemma}\label{lem:coefficient-separation}
Let $b\in R_n$ satisfy $\norm b_\infty\le1$. For every $0\le j<m$, $(\polyk b)_j=0$ if and only if $(\polyazero b)_j=(\polyaone b)_j=0$.
\end{lemma}
\begin{proof}
If $(\polyazero b)_j=(\polyaone b)_j=0$, then $(\polyk b)_j=0$ by the definition of $\polyk$. Conversely, suppose that $(\polyk b)_j=0$. The identity $(\polyazero b)_j=-\packbase(\polyaone b)_j$ shows that both the real and imaginary parts of $(\polyazero b)_j$ are integers divisible by $\packbase$. Since $\norm b_\infty\le1$, inequality~\eqref{eq:product-bound} gives
\[
 \norm{\polyazero b}_\infty\le\norm{\polyazero}_1\norm b_\infty
 \le2(m-1)<\packbase.
\]
Both parts must therefore be zero, and hence $(\polyazero b)_j=(\polyaone b)_j=0$.\qed
\end{proof}

We next consider $b\in\GU$, or equivalently $\norm{\polyd b}_\infty=1$ by Lemma~\ref{lem:gaussian-unit}. Lemmas~\ref{lem:two-moments} and~\ref{lem:coefficient-separation} give the following criterion for signed monomials.

\begin{corollary}\label{cor:packed-moments}
Let $b\in\GU$. If $(\polyk b)_j=0$ for every $0\le j<m$ with $b_j\ne0$, then $b=\pm X^\ell$ for some $0\le\ell<2m$.
\end{corollary}
\begin{proof}
Since $b\in\GU$, we have $\norm b_\infty=1$. Lemmas~\ref{lem:two-moments} and~\ref{lem:coefficient-separation} therefore show that $b$ has exactly one nonzero complex coefficient. This coefficient belongs to $\{\pm1,\pm\ii\}$, so $\ii=X^m$ gives $b=\pm X^\ell$ for some $0\le\ell<2m$.\qed
\end{proof}

Equation~\eqref{eq:two-moments} gives $(\polyazero)^*=-\polyazero$ and $(\polyaone)^*=-\polyaone$. By linearity, $\polyk^*=-\polyk$. For the construction in Section~\ref{subsec:module-construction}, we also need the following coefficient identities and norm bounds for $\polyk$ and $\polyd\polyk$.

\begin{lemma}\label{lem:module-estimates}
The polynomial $\polyk$ satisfies
\begin{enumerate}
\item $[\polyk]_0=[\polyk]_m=[\polyd\polyk]_0=[\polyd\polyk]_m=0$.
\item $\norm{\polyd\polyk}_\infty\le\norm{\polyk}_1<3m^3$.
\end{enumerate}
\end{lemma}
\begin{proof}
Using $X^{-\delta}=-X^{n-\delta}$ in~\eqref{eq:packed-moments}, we find that the nonzero ordinary coefficients of $\polyk$ occur at indices $1,\ldots,m-1$ and $m+1,\ldots,n-1$. Thus $[\polyk]_0=[\polyk]_m=0$, and the coefficient calculation in Lemma~\ref{lem:gaussian-unit} gives
\[
 ([\polyd\polyk]_0,[\polyd\polyk]_m)
 =([\polyk]_0-[\polyk]_m,[\polyk]_0+[\polyk]_m)=(0,0).
\]
Since $\norm{\polyd}_\infty=1$, inequality~\eqref{eq:product-bound} and the distinct ordinary coefficient positions in~\eqref{eq:packed-moments} give
\[
 \norm{\polyd\polyk}_\infty\le \norm{\polyk}_1
 =2\sum_{\delta=1}^{m-1}(1+\packbase\delta)
 =(m-1)(2+\packbase m)<3m^3.\tag*{\(\qed\)}
\]
\end{proof}

\subsection{The rank-two module construction}\label{subsec:module-construction}
We now construct an instance  $(n,\mathcal M,M)$ of $\MSVP_{\infty,2}$ over the ring $R_n=\Z[X]/(X^n+1)$ from a \SSum\ instance $(w_1,\ldots,w_s;B)$ with $s\ge1$ positive integer weights and a nonnegative integer target $B$. Set $W\coloneqq\sum_{j=1}^s w_j$. If $B>W$, the \SSum\ instance is a NO instance. In this case,  we set $\mathcal M\coloneqq2R_2^2$ with basis $2\matI_2$ and output $(2,\mathcal M,1)$.

We assume $0\le B\le W$ from now on. We choose the least power of two $m>s$ and set $n\coloneqq2m$.

\par\medskip\noindent\textbf{Step 1. Encoding the instance.} We first specify the positions of the Boolean variables using the polynomial $\polyu\coloneqq\sum_{j=1}^sX^{j-1}$, whose coefficients are $1$ at indices $0,\ldots,s-1$ and $0$ elsewhere. To obtain the weighted sum in the constant coefficient, we define $\polyhbase\coloneqq\sum_{j=1}^s w_jX^{-(j-1)}$, so that $[\polyhbase a]_0=\sum_{j=1}^s w_j[a]_{j-1}$ for $a\in R_n$. We then set
\begin{equation}\label{eq:module-source}
 \polyh\coloneqq(1-X^m)\polyhbase,\qquad
 \polyv\coloneqq(2B-W)(1-X^m).
\end{equation}
For $b=1$ and $a=\sum_{j=1}^s\xi_jX^{j-1}$ with $\xi_j\in\{0,1\}$ and $z=2a-\polyu$, the constant coefficient and the coefficient of $X^m$ in $\polyh z-\polyv$ are $2(\sum_{j=1}^s w_j\xi_j-B)$ and its negative, respectively.

\par\medskip\noindent\textbf{Step 2. Constructing the module.} Since $0\le B\le W$, we have $\norm{\polyh}_1\le2\norm{\polyhbase}_1=2W$ and $\norm{\polyv}_1=2|2B-W|\le2W$. Recall that $\polyd=1+X^m$. We choose $L$ and $M$ and define $\polyq$ by
\begin{equation}\label{eq:module-scales}
 L\coloneqq4W+1,\qquad M\coloneqq2L{\norm{\polyk}_1}+6W+1,\qquad
 \polyq\coloneqq M\polyd+L\polyd\polyk.
\end{equation}

We define the generator matrix
\begin{equation}\label{eq:module-generators}
 \matG\coloneqq\begin{pmatrix}
 2\polyh&-\polyh\polyu-\polyq-\polyv\\
 2M&-M\polyu
 \end{pmatrix}\in R_n^{2\times2}.
\end{equation}
Let $\mathcal M\coloneqq\matG R_n^2$. With $z\coloneqq2a-\polyu b$ for $a,b\in R_n$, we have
\begin{equation}\label{eq:module-vector}
 \matG\binom ab
 =\binom{\polyh z-(\polyq+\polyv)b}{Mz}.
\end{equation}
We output the instance  $(n,\mathcal M,M)$.

\subsection{Proof of the reduction}\label{subsec:module-correctness}

We first verify the signed-monomial restriction and compute the exact norm for Boolean choices. We then establish the correctness and polynomial running time of the reduction in Section~\ref{subsec:module-construction}.

\begin{lemma}\label{lem:module-monomial}
Let $\polyh,\polyv,L,M,\polyq$ be defined by~\eqref{eq:module-source} and~\eqref{eq:module-scales}, with $0\le B\le W$. Let $b,z\in R_n$ satisfy $\norm z_\infty\le1$ and
\begin{equation}\label{eq:module-monomial-bound}
 \norm{\polyh z-(\polyq+\polyv)b}_\infty\le M.
\end{equation}
Then either $b=0$ or $b=\pm X^\ell$ for some $0\le\ell<n$.
\end{lemma}
\begin{proof}
Assume $b\ne0$. To apply Corollary~\ref{cor:packed-moments}, we show that $b\in\GU$ and $(\polyk b)_j=0$ for every $j$ with $b_j\ne0$. Set $e\coloneqq\polyv b-\polyh z$ and $t\coloneqq\norm{\polyd b}_\infty$.

By Lemma~\ref{lem:gaussian-unit}, the condition $b\in\GU$ is equivalent to $t=1$. We first compare $M\polyd b$ with the possible cancellation from $L\polyd\polyk b+e$ to rule out $t\ge2$. Equation~\eqref{eq:D-norms} gives $1\le\norm b_\infty\le t$. Using~\eqref{eq:product-bound} and the bounds $\norm{\polyh}_1,\norm{\polyv}_1\le2W$, we obtain
\[
 \norm e_\infty\le2W(t+1),\qquad
 \norm{\polyd\polyk b}_\infty\le {\norm{\polyk}_1}t.
\]
By~\eqref{eq:module-monomial-bound} and the definition of $\polyq$, we have $\norm{M\polyd b+L\polyd\polyk b+e}_\infty\le M$. If $t\ge2$, the triangle inequality gives
\[
 \begin{aligned}
 \norm{M\polyd b+L\polyd\polyk b+e}_\infty
 &\ge Mt-L\norm{\polyd\polyk b}_\infty-\norm e_\infty\\
 &\ge(M-L\norm{\polyk}_1-2W)t-2W\\
 &\ge M+1.
 \end{aligned}
\]
The last inequality follows from $t\ge2$ and the definition of $M$. This contradicts~\eqref{eq:module-monomial-bound}. Since $t$ is a positive integer, we conclude that $t=1$. Lemma~\ref{lem:gaussian-unit} gives $b\in\GU$, and $\norm e_\infty\le4W=L-1$.

We next use $\polyk^*=-\polyk$ to prove that $(\polyk b)_j=0$ for every $j$ with $b_j\ne0$. Fix $0\le\ell<n$ with $[\polyd b]_\ell\ne0$. Since $\norm{\polyd b}_\infty=1$, we have $[\polyd b]_\ell\in\{-1,1\}$. Multiplying the expression inside $|M[\polyd b]_\ell+L[\polyd\polyk b]_\ell+[e]_\ell|\le M$ by $[\polyd b]_\ell$ and using $[\polyd b]_\ell^2=1$ gives
\[
 M+L[\polyd b]_\ell[\polyd\polyk b]_\ell+[\polyd b]_\ell[e]_\ell\le M.
\]
Subtracting $M$ and using $\norm e_\infty\le L-1$ yields
\[
 L[\polyd b]_\ell[\polyd\polyk b]_\ell
 \le-[\polyd b]_\ell[e]_\ell\le|[e]_\ell|\le L-1<L.
\]
By integrality, $[\polyd b]_\ell[\polyd\polyk b]_\ell\le0$. At a position where $[\polyd b]_\ell=0$, the product is zero. Since $\polyk^*=-\polyk$, we also have
\[
 \sum_{\ell=0}^{n-1}[\polyd b]_\ell[\polyd\polyk b]_\ell
 =\ip{\polyd b}{\polyk(\polyd b)}=0.
\]
Every summand is nonpositive, so each is zero. Hence $[\polyd\polyk b]_\ell=0$ for every $\ell$ with $[\polyd b]_\ell\ne0$.

For every $j$ with $b_j\ne0$, we have $b_j\in\{1,-1,\ii,-\ii\}$. Both the real and imaginary parts of $(1+\ii)b_j$ are therefore nonzero, so $[\polyd b]_j$ and $[\polyd b]_{j+m}$ are nonzero. Consequently $[\polyd\polyk b]_j=[\polyd\polyk b]_{j+m}=0$, and hence $(\polyd\polyk b)_j=(1+\ii)(\polyk b)_j=0$. Since $1+\ii\ne0$, we obtain $(\polyk b)_j=0$ for every $j$ with $b_j\ne0$. Corollary~\ref{cor:packed-moments} yields $b=\pm X^\ell$.\qed
\end{proof}

The next lemma computes the norm of the vector in~\eqref{eq:module-vector} for $b=1$ and Boolean choices.

\begin{lemma}\label{lem:module-boolean-norm}
Let $\matG$ be given by~\eqref{eq:module-generators} with the choices in~\eqref{eq:module-source} and~\eqref{eq:module-scales}, where $0\le B\le W$. For $\vecxi\in\{0,1\}^s$, set $a\coloneqq\sum_{j=1}^s\xi_jX^{j-1}$. Then
\begin{equation}\label{eq:module-norm}
 \norm{\matG\binom a1}_\infty
 ={\norm{\binom{\polyh(2a-\polyu)-\polyq-\polyv}{M(2a-\polyu)}}_\infty}
 =M+2\left|\sum_{j=1}^s w_j\xi_j-B\right|.
\end{equation}
\end{lemma}
\begin{proof}
Set $A\coloneqq\sum_{j=1}^s w_j\xi_j$ and $z\coloneqq2a-\polyu=\sum_{j=1}^s(2\xi_j-1)X^{j-1}$. Then $\norm z_\infty=1$. Expanding gives
\[
 \polyhbase z
 =\sum_{j,\ell=1}^s w_j(2\xi_\ell-1)X^{\ell-j}.
\]
Every exponent lies in $[-s+1,s-1]\subseteq[-m+1,m-1]$. Thus only the terms with $j=\ell$ contribute to the constant coefficient, and the coefficient of $X^m$ is zero. Hence
\[
 [\polyhbase z]_0=\sum_{j=1}^s w_j(2\xi_j-1)=2A-W,\qquad
 [\polyhbase z]_m=0.
\]
From $\polyh=(1-X^m)\polyhbase$ and $\polyv=(2B-W)(1-X^m)$, we obtain $[\polyh z-\polyv]_0=2(A-B)$ and $[\polyh z-\polyv]_m=-2(A-B)$. Also, $\polyq=M\polyd+L\polyd\polyk$ and Lemma~\ref{lem:module-estimates}(i) give $[\polyq]_0=[\polyq]_m=M$. Therefore
\begin{equation}\label{eq:module-residual}
 \begin{aligned}
 [\polyh z-\polyq-\polyv]_0&=-M+2(A-B),\\
 [\polyh z-\polyq-\polyv]_m&=-M-2(A-B).
 \end{aligned}
\end{equation}
The two coefficients in~\eqref{eq:module-residual} satisfy
\[
 \max\bigl\{|-M+2(A-B)|,\ |-M-2(A-B)|\bigr\}
 =M+2|A-B|\ge M.
\]

At every position $\ell\notin\{0,m\}$, the coefficients of $M\polyd$ and $\polyv$ are zero. Since $\norm{\polyh z}_\infty\le2W$ by~\eqref{eq:product-bound}, Lemma~\ref{lem:module-estimates}(ii) gives
\[
 |[\polyh z-\polyq-\polyv]_\ell|
 \le\norm{\polyh z}_\infty+L\norm{\polyd\polyk}_\infty
 \le2W+L{\norm{\polyk}_1}<M.
\]
Thus $\norm{\polyh z-\polyq-\polyv}_\infty=M+2|A-B|$. Since $\norm{Mz}_\infty=M$, equation~\eqref{eq:module-vector} gives \eqref{eq:module-norm}.\qed
\end{proof}

We now verify the reduction.

\begin{theorem}\label{thm:module}
There exists a deterministic polynomial-time reduction from \SSum\ to $\MSVP_{\infty,2}$ that maps each instance to an instance  $(n,\mathcal M,M)$, where  $\mathcal M=\matG R_n^2$ is a free rank-two module lattice over $R_n$. The first minimum of  $\mathcal M$ is $M$ on YES instances and at least $M+1$ on NO instances. Consequently, $\MSVP_{\infty,2}$ is $\NP$-complete for free rank-two modules over $R_n$, where $n$ ranges over powers of two.
\end{theorem}

\begin{proof}
For $B>W$,  the output module $\mathcal M=2R_2^2$ has first minimum $2$. Assume $0\le B\le W$. By Lemma~\ref{lem:module-estimates}(i), $[\polyq+\polyv]_0=M+2B-W\ge M-W>0$. Since $R_n$ is an integral domain, $\det\matG=2M(\polyq+\polyv)\ne0$, so $\matG$ defines  a free module of rank two.

Every solution to the \SSum\ instance gives a nonzero vector of norm $M$ by Lemma~\ref{lem:module-boolean-norm}. Conversely, let $\matG\binom ab$ be a nonzero vector of norm at most $M$, and put $z\coloneqq2a-\polyu b$. Equation~\eqref{eq:module-vector} gives $\norm z_\infty\le1$ and $\norm{\polyh z-(\polyq+\polyv)b}_\infty\le M$, so Lemma~\ref{lem:module-monomial} gives $b=0$ or $b=\pm X^\ell$. The case $b=0$ would force $\norm{2a}_\infty\le1$ and hence $a=0$, contrary to the vector being nonzero. Thus $b=\pm X^\ell$. Multiplying the vector by $b^{-1}$ preserves its coefficient norm, so we may replace $a$ by $ab^{-1}$ and assume $b=1$.

Now $\norm{2a-\polyu}_\infty\le1$ forces $[a]_j\in\{0,1\}$ for $0\le j<s$ and $[a]_j=0$ for $s\le j<n$. Hence
\[
 a=\sum_{j=1}^s\xi_jX^{j-1},
\]
where $\vecxi\in\{0,1\}^s$. Lemma~\ref{lem:module-boolean-norm} gives $M+2|\sum_{j=1}^s w_j\xi_j-B|\le M$, forcing $\sum_{j=1}^s w_j\xi_j=B$ and the vector norm to equal $M$. Therefore the first minimum is $M$ on a YES instance and at least $M+1$ on a NO instance by integrality.

The choice of $m$ gives $m\le2s$, so the integer lattice dimension is $2n\le8s$. Since $L=O(W)$ and $\norm{\polyk}_1=O(m^3)$, we have $M=O(m^3W)=O(s^3W)$. Also, $\norm{\polyh\polyu}_\infty\le2W$ and
\[
 \norm{-\polyh\polyu-\polyq-\polyv}_\infty
 \le2W+M+L{\norm{\polyk}_1}+2W<2M.
\]
The coefficients of the other entries $2\polyh$, $-M\polyu$, and $2M$ also have absolute value at most $2M$. Thus the output has $O(s(\log s+\log W))$ bits, and the explicit formulas give a deterministic polynomial-time construction. Membership in $\NP$ follows from Lemma~\ref{lem:integer-operations}.\qed
\end{proof}

\begin{samepage}
The proof of Theorem~\ref{thm:module} also gives the following corollary. Comparing the norm of a shortest vector with $M$ decides the source \SSum\ instance, and on a YES instance the proof recovers a solution in polynomial time.

\begin{corollary}\label{cor:module-search}
Search-$\MSVP_{\infty,2}$ is $\NP$-hard for free rank-two modules over $R_n$ under polynomial-time Turing reductions, where $n$ ranges over powers of two.
\end{corollary}
\end{samepage}
% End the local notation settings for Section 3.

% Confirmed Section 4: future edits remain marked with {...}.
\section{Principal ideals}\label{sec:principal}

In this section, we encode a \SSum\ instance $(w_1,\ldots,w_s;B)$ in a principal ideal $(f)\subseteq R_n$ whose nonzero vectors of $\ell_\infty$-norm at most $\radius$ determine solutions to the instance. We choose a generator of the form $f=(\radius+C-\polyg)\polyd$, where $\polyd=1+X^m$. The polynomial $\polyg$ and the positive integers $\radius,C$ will be specified in Section~\ref{subsec:principal-construction}.

We choose $\radius$ so that every nonzero vector $fb$ satisfying $\norm{fb}_\infty\le\radius$ has $\norm{\polyd b}_\infty=1$, and hence $b\in\GU$ by Lemma~\ref{lem:gaussian-unit}. The indices $j$ with $b_j\ne0$ encode the Boolean choices. We construct $\polyg$ so that all nonzero complex coefficients of $b$ become $1$ after multiplication by a power of $X$. Their indices then consist of $0$ and one index in each of $s$ separated three-element sets. The position chosen in each set gives a Boolean value $\xi_h\in\{0,1\}$. The bounds on the ordinary coefficients at indices $0$ and $m$ then force $\sum_{h=1}^s w_h\xi_h=B$.

\subsection[Norm identities and support restrictions]{Norm identities and support restrictions}\label{subsec:principal-tools}
Let $m\ge2$ be a power of two and set $n\coloneqq2m$. Recall that $\polyd=1+\ii=1+X^m$, and use the complex coefficients in~\eqref{eq:gaussian-blocks}. We first prove the norm identities with a positive integer $M$ as the main-term coefficient. The construction in Section~\ref{subsec:principal-construction} takes $M=\radius+C$. For $\polyg\in R_n$, define
\begin{equation}\label{eq:principal-conversion}
 f\coloneqq(M-\polyg)\polyd.
\end{equation}
We first compare $\norm{fb}_\infty$ for arbitrary nonzero $b\in R_n$. The triangle inequality and~\eqref{eq:product-bound} give
\begin{equation}\label{eq:dominant-operator}
 (M-\norm{\polyg}_1)\norm{\polyd b}_\infty
 \le\norm{fb}_\infty
 \le(M+\norm{\polyg}_1)\norm{\polyd b}_\infty.
\end{equation}
By Lemma~\ref{lem:gaussian-unit}, $\norm{\polyd b}_\infty=1$ for $b\in\GU$, whereas $\norm{\polyd b}_\infty\ge2$ for every other nonzero $b$. In particular, $\norm{f\sigma}_\infty\le M+\norm{\polyg}_1$ for every $\sigma\in\GU$. We assume $M>3\norm{\polyg}_1$, so that $2(M-\norm{\polyg}_1)>M+\norm{\polyg}_1$. Thus $f\ne0$, and every shortest nonzero vector of $(f)$ has the form $f\sigma$ with $\sigma\in\GU$.

It therefore suffices to determine the coefficient $\ell_\infty$-norm of $f\sigma$ for $\sigma\in\GU$. If $\sigma_j\ne0$, both $[\polyd\sigma]_j$ and $[\polyd\sigma]_{j+m}$ belong to $\{-1,1\}$, so the corresponding coefficients of $M\polyd\sigma$ have absolute value $M$. For $\ell\in\{j,j+m\}$, multiplying inside the absolute value by $[\polyd\sigma]_\ell$ gives
\[
 |[f\sigma]_\ell|
 =\bigl|M-[\polyd\sigma]_\ell[\polyd\polyg\sigma]_\ell\bigr|.
\]\par

For $0\le j<m$ with $\sigma_j\ne0$, define
\begin{equation}\label{eq:local-values}
 \phi_j^\pm(\polyg,\sigma)\coloneqq
 \Re\bigl(\overline{\sigma_j}(\polyg\sigma)_j\bigr)
 \pm\Im\bigl(\overline{\sigma_j}(\polyg\sigma)_j\bigr).
\end{equation}
If $\overline{\sigma_j}(\polyg\sigma)_j=x+\ii y$, then the minimum decrease in absolute value of the two ordinary coefficients at $j$ and $j+m$ is $\min\{x-y,x+y\}=x-|y|$ by Lemma~\ref{lem:signed-values}. For fixed $\sigma$ and $j$, the maps $\polyg\mapsto \phi_j^\pm(\polyg,\sigma)$ take values in $\Z$ and are $\Z$-linear. In particular, $\phi_j^\pm(c\polyg+\polyt,\sigma) =c\phi_j^\pm(\polyg,\sigma)+\phi_j^\pm(\polyt,\sigma)$ for $c\in\Z$ and $\polyg,\polyt\in R_n$. The following lemma expresses the absolute values of the ordinary coefficients of $f\sigma$ at $j$ and $j+m$ as $|M-\phi_j^\pm(\polyg,\sigma)|$.

\begin{lemma}\label{lem:signed-values}
Let $\polyg\in R_n$ and $\sigma\in\GU$. For every $0\le j<m$ with $\sigma_j\ne0$,
\[
 \bigl\{[\polyd\sigma]_j[\polyd\polyg\sigma]_j,\,
        [\polyd\sigma]_{j+m}[\polyd\polyg\sigma]_{j+m}\bigr\}
 =\bigl\{\phi_j^-(\polyg,\sigma),\,\phi_j^+(\polyg,\sigma)\bigr\}.
\]
Moreover, $|\phi_j^\pm(\polyg,\sigma)|\le\norm{\polyg}_1$.
\end{lemma}
\begin{proof}
Fix $0\le j<m$ with $\sigma_j\ne0$ and write $\overline{\sigma_j}(\polyg\sigma)_j=x+\ii y$ with $x,y\in\R$. By~\eqref{eq:local-values}, $\phi_j^-(\polyg,\sigma)=x-y$ and $\phi_j^+(\polyg,\sigma)=x+y$. Since $\sigma_j\overline{\sigma_j}=1$, we have $(\polyg\sigma)_j=\sigma_j(x+\ii y)$. Multiplication by $\polyd=1+\ii$ gives
\[
 \polyd(x+\ii y)=x+\ii y+\ii x-y=(x-y)+\ii(x+y).
\]
Hence $(\polyd\sigma)_j=\sigma_j(1+\ii)$ and $(\polyd\polyg\sigma)_j=\sigma_j\bigl((x-y)+\ii(x+y)\bigr)$. Computing the ordinary coefficient products yields
\[
 \bigl([\polyd\sigma]_j[\polyd\polyg\sigma]_j,\,
       [\polyd\sigma]_{j+m}[\polyd\polyg\sigma]_{j+m}\bigr)
 =\begin{cases}
   (x-y,x+y),&\sigma_j\in\{1,-1\},\\
   (x+y,x-y),&\sigma_j\in\{\ii,-\ii\}.
  \end{cases}
\]
In both cases, the unordered pair is $\{\phi_j^-(\polyg,\sigma),\phi_j^+(\polyg,\sigma)\}$, proving the identity.

By Lemma~\ref{lem:gaussian-unit}, $\norm{\polyd\sigma}_\infty=1$. Inequality~\eqref{eq:product-bound} therefore gives
\[
 \norm{\polyd\polyg\sigma}_\infty
 \le\norm{\polyg}_1\norm{\polyd\sigma}_\infty
 =\norm{\polyg}_1.
\]
Since $[\polyd\sigma]_j,[\polyd\sigma]_{j+m}\in\{-1,1\}$, this also bounds $|\phi_j^\pm(\polyg,\sigma)|$.\qed
\end{proof}

For $\sigma\in\GU$, define the minimum of $\phi_j^\pm(\polyg,\sigma)$ over all indices $j$ with $\sigma_j\ne0$ by
\[
 \theta_{\polyg}(\sigma)\coloneqq
 \min_{\substack{0\le j<m\\\sigma_j\ne0}}
 \min\{\phi_j^+(\polyg,\sigma),\phi_j^-(\polyg,\sigma)\}.
\]
The following lemma shows how $\theta_{\polyg}(\sigma)$ determines the coefficient $\ell_\infty$-norm of $f\sigma$.

\begin{lemma}\label{thm:gaussian-conversion}
Assume $M>3\norm{\polyg}_1$, and let $f$ be defined by~\eqref{eq:principal-conversion}. For every $\sigma\in\GU$, we have $\norm{f\sigma}_\infty=M-\theta_{\polyg}(\sigma)$.
\end{lemma}
\begin{proof}
We compute the absolute values of the ordinary coefficients of $f\sigma$ for $\sigma\in\GU$. Fix $0\le j<m$. If $\sigma_j\ne0$, equation~\eqref{eq:principal-conversion} and Lemma~\ref{lem:signed-values} give $|M-\phi_j^\pm(\polyg,\sigma)|$ as the two absolute values at indices $j$ and $j+m$. Since $|\phi_j^\pm(\polyg,\sigma)|\le\norm{\polyg}_1<M$, these equal $M-\phi_j^\pm(\polyg,\sigma)\ge M-\norm{\polyg}_1$. If $\sigma_j=0$, both ${[\polyd\sigma]_j}$ and ${[\polyd\sigma]_{j+m}}$ are zero, so for $\ell\in\{j,j+m\}$,
\[
 |[f\sigma]_\ell|
 =|[\polyd\polyg\sigma]_\ell|
 \le\norm{\polyg}_1<M-\norm{\polyg}_1.
\]
Taking the maximum over all ordinary coefficients therefore gives $\norm{f\sigma}_\infty=M-\theta_{\polyg}(\sigma)$.\qed
\end{proof}

For $\sigma\in\GU$, Lemma~\ref{thm:gaussian-conversion} shows that $\norm{f\sigma}_\infty\le M$ if and only if $\theta_{\polyg}(\sigma)\ge0$. Since multiplication by $X^\ell$ permutes the ordinary coefficients of $f\sigma$ up to sign, $\norm{f(X^\ell\sigma)}_\infty=\norm{f\sigma}_\infty$ for every integer $\ell$. The norm bound therefore cannot distinguish a common cyclic translation of the indices of nonzero complex coefficients. We choose $\polyg$ so that $\theta_{\polyg}(\sigma)\ge0$ restricts the cyclic distances between these indices.

The \emph{cyclic distance} between $j,\ell\in\{0,\ldots,m-1\}$ is defined as $\min\{|j-\ell|,m-|j-\ell|\}$. This counts the steps along the shorter of the two paths between $j$ and $\ell$ around the cycle. Let $E\subseteq\{1,\ldots,m/2\}$ be the set of allowed nonzero cyclic distances, and order the elements of $\{1,\ldots,m/2\}\setminus E$ as $\Delta_0<\Delta_1<\cdots<\Delta_{r-1}$, where $r\coloneqq m/2-|E|$.

To exclude a cyclic distance $\delta\notin E$ between indices of nonzero complex coefficients, we require $\sigma_{j-\delta}-\sigma_{j+\delta}=0$ for every $j$ with $\sigma_j\ne0$. As we will show in Lemma~\ref{lem:gaussian-distances}, if $\sigma_j$ and $\sigma_{j+\delta}$ were both nonzero, these equalities would force $(\sigma_{j+\nu\delta})_{\nu\in\Z}$ to be nonzero and $2$-periodic. For power-of-two $m$, this contradicts~\eqref{eq:gaussian-extension}. The difference $\sigma_{j-\delta}-\sigma_{j+\delta}$ is the $j$-th complex coefficient of $(X^\delta-X^{-\delta})\sigma$. We combine these differences over the distances outside $E$ and define
\begin{equation}\label{eq:forbidden-polynomial}
 \polyae\coloneqq\sum_{\nu=0}^{r-1}5^\nu(X^{\Delta_\nu}-X^{-\Delta_\nu}).
\end{equation}
The powers of $5$ separate the coefficient conditions for different distances. For $\sigma\in\GU$, requiring $(\polyae\sigma)_j=0$ at every index $j$ with $\sigma_j\ne0$ excludes pairs of nonzero complex coefficients at cyclic distances outside $E$. The next lemma proves this implication and shows how to choose $\polyg$ so that $\theta_{\polyg}(\sigma)\ge0$ forces these zero coefficients.

\begin{lemma}\label{lem:gaussian-distances}
Let $\sigma\in\GU$. Then $(\polyae\sigma)_j=0$ for every $j$ with $\sigma_j\ne0$ if and only if all pairwise cyclic distances between the positions of the nonzero complex coefficients of $\sigma$ belong to $E$. Moreover, let $\polyt\in R_n$, and set $L\coloneqq\norm{\polyt}_1+1$ and $\polyg\coloneqq L\polyae+\polyt$. If $\theta_{\polyg}(\sigma)\ge0$, both conditions hold.
\end{lemma}
\begin{proof}
We first prove the equivalence. If all pairwise cyclic distances between indices of nonzero complex coefficients belong to $E$, then $\sigma_{j-\delta}=\sigma_{j+\delta}=0$ for every $j$ with $\sigma_j\ne0$ and $\delta\in\{1,\ldots,m/2\}\setminus E$, so~\eqref{eq:forbidden-polynomial} gives $(\polyae\sigma)_j=0$.

Conversely, suppose that $(\polyae\sigma)_j=0$ for every $j$ with $\sigma_j\ne0$, and fix $0\le j<m$ with $\sigma_j\ne0$. Using~\eqref{eq:gaussian-extension}, we compute the $j$-th complex coefficient of $\polyae\sigma$ as
\[
 (\polyae\sigma)_j
 =\sum_{\nu=0}^{r-1}5^\nu
       (\sigma_{j-\Delta_\nu}-\sigma_{j+\Delta_\nu})=0.
\]
For every $0\le\nu<r$, the real and imaginary parts of $\sigma_{j-\Delta_\nu}-\sigma_{j+\Delta_\nu}$ lie in $\{-2,-1,0,1,2\}$. If $r>0$, reduction modulo $5$ gives $\sigma_{j-\Delta_0}=\sigma_{j+\Delta_0}$. Dividing by $5$ and repeating the argument gives
\begin{equation}\label{eq:gaussian-recurrence}
 \sigma_{j-\delta}=\sigma_{j+\delta},
\end{equation}
for every $\delta\in\{1,\ldots,m/2\}\setminus E$. By~\eqref{eq:gaussian-extension}, shifting $j$ by $qm$ for any $q\in\Z$ multiplies both sides of~\eqref{eq:gaussian-recurrence} by $\ii^{-q}$. Thus~\eqref{eq:gaussian-recurrence} holds for every $j\in\Z$ with $\sigma_j\ne0$.

Suppose two indices of nonzero complex coefficients have cyclic distance $\delta\notin E$. Choose $j\in\Z$ so that $a_0\coloneqq\sigma_j$ and $a_1\coloneqq\sigma_{j+\delta}$ are nonzero, and define $a_\nu\coloneqq\sigma_{j+\nu\delta}$ for every $\nu\in\Z$. If $a_\nu\ne0$, equation~\eqref{eq:gaussian-recurrence} gives $a_{\nu-1}=a_{\nu+1}$. Starting from $a_0,a_1\ne0$ and inducting in both directions shows that every $a_\nu$ is nonzero and $a_{\nu+2}=a_\nu$.

Set $\eta\coloneqq\gcd(m,\delta)$, $\ell\coloneqq m/\eta$, and $q\coloneqq\delta/\eta$. Since $m$ is a power of two and $1\le\delta\le m/2$, $\ell$ is even and $q$ is odd. Since $\ell\delta=qm$, the $2$-periodicity and~\eqref{eq:gaussian-extension} give
\[
 a_0=a_\ell=\sigma_{j+\ell\delta}
 =\sigma_{j+qm}
 =\ii^{-q}a_0.
\]
Since $q$ is odd, $\ii^{-q}\in\{\ii,-\ii\}$. The equality $a_0=\ii^{-q}a_0$ therefore implies $a_0=0$, contrary to the choice $a_0=\sigma_j\ne0$. Thus all pairwise cyclic distances between the indices of nonzero complex coefficients belong to $E$.

Finally, assume that $\theta_{\polyg}(\sigma)\ge0$. For every $0\le j<m$ with $\sigma_j\ne0$, linearity and Lemma~\ref{lem:signed-values} give
\[
 0\le\phi_j^\pm(\polyg,\sigma)
 =L\phi_j^\pm(\polyae,\sigma)+\phi_j^\pm(\polyt,\sigma)
 \le L\phi_j^\pm(\polyae,\sigma)+L-1.
\]
Hence $\phi_j^\pm(\polyae,\sigma)>-1$, and integrality gives $\phi_j^\pm(\polyae,\sigma)\ge0$.

By linearity and~\eqref{eq:forbidden-polynomial}, $\polyae^*=-\polyae$. Using~\eqref{eq:adjoint-bound}, the complex coefficient representation, and~\eqref{eq:local-values}, we obtain
\begin{equation}\label{eq:ae-orthogonality}
 \begin{aligned}
 \sum_{\substack{0\le j<m\\\sigma_j\ne0}}
 \bigl(\phi_j^+(\polyae,\sigma)+\phi_j^-(\polyae,\sigma)\bigr)
 &=2\Re\sum_{j=0}^{m-1}\overline{\sigma_j}(\polyae\sigma)_j\\
 &=2\sum_{\ell=0}^{2m-1}[\sigma]_\ell[\polyae\sigma]_\ell
 =2\ip{\sigma}{\polyae\sigma}=0.
 \end{aligned}
\end{equation}
Since each $\phi_j^\pm(\polyae,\sigma)$ is nonnegative, equation~\eqref{eq:ae-orthogonality} gives
\[
 \phi_j^+(\polyae,\sigma)=\phi_j^-(\polyae,\sigma)=0
\]
for every $j$ with $\sigma_j\ne0$. Taking the sum and difference in~\eqref{eq:local-values} gives $\overline{\sigma_j}(\polyae\sigma)_j=0$. Since $\sigma_j\ne0$, we obtain $(\polyae\sigma)_j=0$. The distance condition follows from the equivalence.\qed
\end{proof}

\subsection{The principal-ideal construction}\label{subsec:principal-construction}
We now construct an instance $(n,(f),\radius)$ of $\ISVP_\infty$ over $R_n=\Z[X]/(X^n+1)$ from a given \SSum\ instance. Let $(w_1,\ldots,w_s;B)$ have $s\ge1$ positive integer weights and a nonnegative integer target $B$, and set $W\coloneqq\sum_{h=1}^s w_h$.

\par\medskip\noindent\textbf{Step 1. Encoding Boolean choices.} We encode the Boolean choices in the indices of the nonzero complex coefficients of $\sigma$, using $s$ three-element sets. Set $V\coloneqq8s+1$ and $m\coloneqq\min\{2^\ell:2^\ell>4V\}$, and define
\begin{equation}\label{eq:windows}
 E\coloneqq\bigcup_{h=1}^s\{8h-1,8h,8h+1\}.
\end{equation}
Here $V=\max E$ is the largest allowed distance. If the indices $j$ with $\sigma_j\ne0$ for $\sigma\in\GU$ have pairwise cyclic distances in $E$, Lemma~\ref{lem:windows} applies. After a common cyclic translation modulo $m$, each three-element set then contains at most one such index. Set $n\coloneqq2m$ and form $\polyae$ by~\eqref{eq:forbidden-polynomial}.

\par\medskip\noindent\textbf{Step 2. Assigning the weights.} Within each three-element set $\{8h-1,8h,8h+1\}$, a nonzero complex coefficient at $8h$ encodes $\xi_h=0$ and does not include $w_h$ in the subset, while one at $8h\pm1$ encodes $\xi_h=1$ and includes $w_h$. Set $H\coloneqq W+1$. We assign weight $H$ to position $8h$ and weight $H+w_h$ to each of the positions $8h\pm1$. The shifts $X^{-8h}$ and $X^{-(8h\pm1)}$ bring the corresponding complex coefficients to index $0$. If the $h$-th set contains no nonzero complex coefficient, its base weight $H$ is missing. Lemma~\ref{lem:principal-support} uses $H>W$ to rule out empty sets. We define
\[
 \polyres\coloneqq\ii\Bigl(
 \sum_{h=1}^s\left[
 HX^{-8h}+(H+w_h)(X^{-(8h-1)}+X^{-(8h+1)})\right]-(sH+B)\Bigr).
\]
We use the factor $\ii$ to place the subset-sum difference in the imaginary part of $(\polyres\sigma)_0$. Multiplication by $\polyd=1+\ii$ makes this difference contribute with opposite signs to the ordinary coefficients at indices $0$ and $m$. We will set their common baseline at the norm bound, so a nonzero difference forces at least one absolute value above that bound.

The $3s+1$ integer exponents $0,-8h,-(8h\pm1)$, $1\le h\le s$, are distinct and lie in $[-V,0]$. Since $V<m/4<2m$, these exponents remain distinct modulo $2m$, so their monomials do not combine in $R_n$. Since multiplication by $\ii$ permutes ordinary coefficients up to sign, $\norm{\polyres}_1=4sH+2W+B$. We choose $C\coloneqq\norm{\polyres}_1+1$. By Lemma~\ref{lem:signed-values}, this choice gives $|\phi_j^\pm(\polyres,\sigma)|\le C-1$ for $\sigma\in\GU$ and $0\le j<m$ with $\sigma_j\ne0$.

Once the indices of nonzero complex coefficients of $\sigma$ have pairwise cyclic distances in $E$ and every three-element set contains exactly one index $j$ with $\sigma_j\ne0$, Lemma~\ref{lem:windows} gives gaps in $\{7,8,9\}$ between consecutive indices of nonzero complex coefficients in $[0,V]$. After normalizing $\sigma_0$ to $1$, we force every other nonzero complex coefficient to equal $1$ by comparing coefficients at consecutive nonzero positions. We therefore define $\polytzero\coloneqq C\sum_{\nu=7}^{9}(2X^\nu+X^{-\nu})$. At a nonzero position between the endpoints, multiplication by $\polytzero$ combines the preceding and following complex coefficients with weights $2C$ and $C$, respectively. With $\theta_{\polyg}(\sigma)\ge C$, the weights $2C$ and $C$ force consecutive nonzero complex coefficients of $\sigma$ to agree, as proved in Lemma~\ref{lem:principal-support}.

\par\medskip\noindent\textbf{Step 3. Constructing the ideal.} The nonzero ordinary coefficients of $\polytzero$ have indices in $[7,9]\cup[2m-9,2m-7]$, and those of $\polyres$ have indices in $[m-V,m]$. These index sets do not intersect because $9\le V<m/4$. Hence $\norm{\polytzero+\polyres}_1=9C+(C-1)=10C-1$. To apply Lemma~\ref{lem:gaussian-distances}, define
\begin{equation}\label{eq:principal-encoding}
 \polyg\coloneqq10C\polyae+\polytzero+\polyres.
\end{equation}

With $\polyd=1+\ii=1+X^m$, choose the norm bound $\radius\coloneqq3\norm{\polyg}_1+1$ and set $M\coloneqq\radius+C$. We define
\begin{equation}\label{eq:principal-output}
 f\coloneqq(\radius+C-\polyg)\polyd.
\end{equation}
Suppose that $\sigma$ has complex coefficient $1$ at $0$ and at one index in each three-element set, and zero elsewhere. Set $Z\coloneqq\sum_{h=1}^s w_h\xi_h-B$. Then $(\polyg\sigma)_0=C+\ii Z$, so the ordinary coefficients of $f\sigma$ at $0$ and $m$ are $\radius+Z$ and $\radius-Z$. A nonzero $Z$ makes one of these coefficients have absolute value greater than $\radius$, as proved in Lemma~\ref{thm:principal-encoding}.

We output the instance $(n,(f),\radius)$ with the integer basis $\matM_f$ from~\eqref{eq:action}.

The next lemma describes the indices allowed by the distance condition for $E$. After a common cyclic translation modulo $m$, the indices outside the origin lie in distinct three-element sets, with all nonzero offsets from the centers $8h$ having the same sign.

\begin{samepage}
\begin{lemma}\label{lem:windows}
Let $J\subseteq\{0,\ldots,m-1\}$ have $\ell\ge1$ elements whose pairwise cyclic distances belong to $E$. There exists $\tau\in\{0,\ldots,m-1\}$ such that, after the cyclic translation $j\mapsto j+\tau\pmod m$, the translated indices have least nonnegative representatives ordered as
\[
 0=p_0<p_1<\cdots<p_{\ell-1}\le V.
\]
Moreover, $p_j=8\kappa_j+\delta_j$ for $1\le j<\ell$, where the $\kappa_j$ are strictly increasing in $\{1,\ldots,s\}$, and either all $\delta_j$ lie in $\{0,1\}$ or all lie in $\{-1,0\}$. In particular, $\ell\le s+1$.
\end{lemma}
\end{samepage}
\begin{proof}
We first apply a cyclic translation that sends one element of $J$ to $0$ modulo $m$, preserving all pairwise cyclic distances. Since $E\subseteq[1,V]$, all indices have representatives in $[-V,V]$. The difference between the largest and smallest representatives is at most $2V<m/2$, so it equals their cyclic distance and is at most $V$. Subtracting the smallest representative gives $0=p_0<\cdots<p_{\ell-1}\le V$.

For $1\le j<\ell$, we have $p_j\le V<m/2$, so the cyclic distance from $0$ to $p_j$ is $p_j$. Each positive $p_j$ belongs to $E$ and has a unique expression $p_j=8\kappa_j+\delta_j$ with $1\le\kappa_j\le s$ and $\delta_j\in\{-1,0,1\}$. Distinct elements of one three-element set have distance at most $2$, whereas $\min E=7$. Hence the positive indices $p_j$ lie in distinct sets, ordered by their centers $8h$, so the $\kappa_j$ are strictly increasing. This gives $\ell\le s+1$, with $\kappa_j=j$ for $1\le j\le s$ when $\ell=s+1$.

For $1\le c<j<\ell$, we have $p_j-p_c=8(\kappa_j-\kappa_c)+(\delta_j-\delta_c)\in E$. Reducing modulo $8$ and using $|\delta_j-\delta_c|\le2$ gives $\delta_j-\delta_c\in\{-1,0,1\}$. Hence $1$ and $-1$ cannot both occur among the $\delta_j$, so all $\delta_j$ lie in $\{0,1\}$ or all lie in $\{-1,0\}$.\qed
\end{proof}

\subsection{Proof of the reduction}\label{subsec:principal-correctness}
For $\sigma\in\GU$, Lemma~\ref{thm:gaussian-conversion} with $M=\radius+C$ gives $\norm{f\sigma}_\infty\le\radius$ if and only if $\theta_{\polyg}(\sigma)\ge C$. Every nonzero $b\in R_n\setminus\GU$ satisfies $\norm{fb}_\infty\ge2(\radius+C-\norm{\polyg}_1)>\radius$ by~\eqref{eq:dominant-operator} and Lemma~\ref{lem:gaussian-unit}, so it suffices to consider $\sigma\in\GU$. We first determine the positions and values of the nonzero complex coefficients of $\sigma$, then test the subset-sum equation.

\begin{lemma}\label{lem:principal-support}
For $\polyg$ constructed in Section~\ref{subsec:principal-construction}, let $\sigma\in\GU$ satisfy $\theta_{\polyg}(\sigma)\ge C$. After multiplication by a power of $X$, $\sigma$ has the form
\[
 \sigma=1+\sum_{h=1}^s X^{8h+\delta_h},
\]
where either all $\delta_h$ lie in $\{0,1\}$ or all lie in $\{-1,0\}$.
\end{lemma}
\begin{proof}
For every integer $\nu$, we have $X^\nu\sigma\in\GU$ and $\norm{f(X^\nu\sigma)}_\infty=\norm{f\sigma}_\infty$, since multiplication by $X^\nu$ permutes ordinary coefficients up to sign. Applying the norm identity in Lemma~\ref{thm:gaussian-conversion} with $M=\radius+C$ to $\sigma$ and $X^\nu\sigma$ gives $\theta_{\polyg}(X^\nu\sigma)=\theta_{\polyg}(\sigma)$.

Since $\theta_{\polyg}(\sigma)\ge C>0$ and $\norm{\polytzero+\polyres}_1=10C-1$, Lemma~\ref{lem:gaussian-distances} with $\polyt=\polytzero+\polyres$ and $L=10C$ shows that all pairwise cyclic distances between the indices of nonzero complex coefficients of $\sigma$ belong to $E$. By Lemma~\ref{lem:windows}, a cyclic translation puts the indices of the nonzero complex coefficients of $\sigma$ in the form $0=p_0<p_1<\cdots<p_{\ell-1}\le V$, with $p_j=8\kappa_j+\delta_j$ for $1\le j<\ell$. Since $\sigma_0\in\{1,-1,\ii,-\ii\}$ and $\ii=X^m$, multiplying further by a power of $X^m$ gives $\sigma_0=1$ and preserves these indices and $\theta_{\polyg}(\sigma)$. Lemma~\ref{lem:gaussian-distances} also gives $(\polyae\sigma)_{p_j}=0$ for $0\le j<\ell$.

We first show that $\ell=s+1$. For $7\le\nu\le9$, we have $m-\nu>V$. Since the nonzero complex coefficients of $\sigma$ have indices in $[0,V]$, \eqref{eq:gaussian-extension} gives $\sigma_{-\nu}=\ii\sigma_{m-\nu}=0$. The definition of $\polytzero$ therefore gives
\[
 (\polytzero\sigma)_0
 =C\sum_{\nu=7}^{9}(2\sigma_{-\nu}+\sigma_\nu)
 =C(\sigma_7+\sigma_8+\sigma_9).
\]
At most one of $\sigma_7,\sigma_8,\sigma_9$ is nonzero, so $\phi_0^\pm(\polytzero,\sigma)\le C$. Together with $(\polyae\sigma)_0=0$ and $\phi_0^+(\polyg,\sigma)\ge C$, linearity in the first argument of $\phi_0^+$ gives $\phi_0^+(\polyres,\sigma)\ge0$.

For $1\le j<\ell$, let $a_j\coloneqq H+w_{\kappa_j}|\delta_j|$ be the weight assigned to the position $p_j$. We have $\Re(\ii\sigma_{p_j})+\Im(\ii\sigma_{p_j})\le1$, and the $\kappa_j$ are distinct. Since $\sigma_0=1$, the definition of $\polyres$ gives
\[
 \phi_0^+(\polyres,\sigma)
 =\sum_{j=1}^{\ell-1}a_j
 \left[\Re(\ii\sigma_{p_j})+\Im(\ii\sigma_{p_j})\right]-sH-B
 \le(\ell-1)H+W-sH-B.
\]
If $\ell\le s$, the upper bound is at most $-H+W-B=-B-1<0$, contradicting $\phi_0^+(\polyres,\sigma)\ge0$. Thus $\ell=s+1$ and $\kappa_j=j$ for $1\le j\le s$.

We next show that all nonzero complex coefficients of $\sigma$ equal $1$. By Lemma~\ref{lem:windows}, $p_j-p_{j-1}\in\{7,8,9\}$ for $1\le j\le s$, whereas $p_j-p_c\ge15$ when $0\le c<j\le s$ and $j-c\ge2$. The definition of $\polytzero$ therefore gives
\[
 (\polytzero\sigma)_{p_j}=
 \begin{cases}
 C\sigma_{p_1},&j=0,\\
 2C\sigma_{p_{j-1}}+C\sigma_{p_{j+1}},&1\le j<s,\\
 2C\sigma_{p_{s-1}},&j=s.
 \end{cases}
\]
Suppose $\sigma_{p_j}\ne\sigma_{p_{j-1}}$ for some $1\le j\le s$, and let $\gamma\coloneqq\sigma_{p_{j-1}}/\sigma_{p_j} =\overline{\sigma_{p_j}}\sigma_{p_{j-1}}\in\{-1,\ii,-\ii\}$ be the ratio of the preceding coefficient to the current one. One of $\Re\gamma\pm\Im\gamma$ equals $-1$. By~\eqref{eq:local-values}, the preceding coefficient contributes $-2C$ to $\phi_{p_j}^\varepsilon(\polytzero,\sigma)$ for some $\varepsilon\in\{+,-\}$. The contribution from the following coefficient is at most $C$ (and is $0$ when $j=s$). Hence $\phi_{p_j}^\varepsilon(\polytzero,\sigma)\le-2C+C$. Lemma~\ref{lem:signed-values} bounds $|\phi_{p_j}^\varepsilon(\polyres,\sigma)|$ by $\norm{\polyres}_1=C-1$. Since $(\polyae\sigma)_{p_j}=0$, we get
\[
 \phi_{p_j}^\varepsilon(\polyg,\sigma)
 \le-2C+C+(C-1)=-1<C,
\]
contradicting $\theta_{\polyg}(\sigma)\ge C$. Thus consecutive coefficients agree. Since $\sigma_{p_0}=1$, all $\sigma_{p_j}$ equal $1$, giving the stated form.\qed
\end{proof}

In the form established in Lemma~\ref{lem:principal-support}, the position of the coefficient $1$ in each three-element set determines a Boolean choice. The next lemma shows that $\norm{f\sigma}_\infty\le\radius$ forces these choices to solve the \SSum\ instance. Conversely, every solution yields a vector of $\ell_\infty$-norm exactly $\radius$.

\begin{lemma}\label{thm:principal-encoding}
For $\polyg$ constructed in Section~\ref{subsec:principal-construction}, we have $\theta_{\polyg}(\sigma)\le C$ for every $\sigma\in\GU$. There exists $\sigma\in\GU$ with $\theta_{\polyg}(\sigma)=C$ if and only if the equation $\sum_{h=1}^s w_h\xi_h=B$ has a solution $\vecxi\in\{0,1\}^s$. Moreover, a solution to the \SSum\ instance can be recovered in deterministic polynomial time from any $\sigma\in\GU$ with $\theta_{\polyg}(\sigma)=C$.
\end{lemma}
\begin{proof}
Suppose $\sigma\in\GU$ satisfies $\theta_{\polyg}(\sigma)\ge C$. By Lemma~\ref{lem:principal-support}, $X^\nu\sigma=1+\sum_{h=1}^s X^{8h+\delta_h}$ for some $\nu\in\Z$. The proof of Lemma~\ref{lem:principal-support} gives $\theta_{\polyg}(X^\nu\sigma)=\theta_{\polyg}(\sigma)$, so we replace $\sigma$ by $X^\nu\sigma$. Set $\xi_h\coloneqq|\delta_h|\in\{0,1\}$ for $1\le h\le s$ and $Z\coloneqq\sum_{h=1}^s w_h\xi_h-B$. At $0$, we have $(\polyae\sigma)_0=0$ by Lemma~\ref{lem:gaussian-distances}, and $(\polytzero\sigma)_0=C$ by the definition of $\polytzero$. Moreover,
\begin{equation}\label{eq:principal-residual}
 (\polyres\sigma)_0
 =\ii\Bigl[\sum_{h=1}^s(H+w_h\xi_h)-(sH+B)\Bigr]
 =\ii Z.
\end{equation}
Since $\sigma_0=1$, we have $\phi_0^\pm(\polyres,\sigma)=\pm Z$. Hence $\phi_0^\pm(\polyg,\sigma)=C\pm Z$. By the definition of $\theta_{\polyg}$,
\[
 C\le\theta_{\polyg}(\sigma)
 \le\min\{C+Z,C-Z\}=C-|Z|.
\]
Hence $Z=0$ and $\theta_{\polyg}(\sigma)=C$, so $\vecxi=(\xi_1,\ldots,\xi_s)$ solves the \SSum\ instance. In particular, $\theta_{\polyg}(\sigma)\le C$ for every $\sigma\in\GU$.

Conversely, let $\vecxi\in\{0,1\}^s$ satisfy $\sum_{h=1}^s w_h\xi_h=B$. Set $p_0\coloneqq0$, $p_h\coloneqq8h+\xi_h$ for $1\le h\le s$, and $\sigma\coloneqq\sum_{h=0}^s X^{p_h}\in\GU$. Every pairwise cyclic distance between its support indices belongs to $E$, so $(\polyae\sigma)_{p_h}=0$ for $0\le h\le s$ by Lemma~\ref{lem:gaussian-distances}. The definition of $\polytzero$ gives $(\polytzero\sigma)_0=C$ and real coefficients $(\polytzero\sigma)_{p_h}\ge2C$ for $1\le h\le s$, since the preceding coefficient $\sigma_{p_{h-1}}=1$ contributes $2C$ and any contribution from the following coefficient is nonnegative. Equation~\eqref{eq:principal-residual} gives $(\polyres\sigma)_0=0$, so $\phi_0^\pm(\polyg,\sigma)=C$. At every other $p_h$, Lemma~\ref{lem:signed-values} gives $\phi_{p_h}^\pm(\polyg,\sigma)\ge2C-(C-1)=C+1$. The minimum defining $\theta_{\polyg}(\sigma)$ is therefore attained at index $0$ and equals $C$. Lemma~\ref{thm:gaussian-conversion} with $M=\radius+C$ gives $\norm{f\sigma}_\infty=\radius$.

Given any $\sigma\in\GU$ with $\theta_{\polyg}(\sigma)=C$, test $X^\nu\sigma$ for $0\le\nu<2n$ to find the form in Lemma~\ref{lem:principal-support}. Replace $\sigma$ by the polynomial found, which still satisfies $\theta_{\polyg}(\sigma)=C$, and let $0=p_0<p_1<\cdots<p_s$ be the indices of its nonzero complex coefficients. Set $\xi_h\coloneqq|p_h-8h|$ for $1\le h\le s$. The inequalities $\phi_0^\pm(\polyg,\sigma)\ge C$ and the coefficient calculation in~\eqref{eq:principal-residual} give $\sum_{h=1}^s w_h\xi_h=B$. The enumeration and recovery take deterministic polynomial time.\qed
\end{proof}

\noindent\begin{minipage}[t]{\linewidth} We now verify the reduction and its running time.

\begin{theorem}\label{thm:principal}
There exists a deterministic polynomial-time reduction from \SSum\ to $\ISVP_\infty$ that maps each instance to an instance of the form $(n,(f),\radius)$, where $0\ne f\in R_n$. The first minimum of $(f)$ is $\radius$ on YES instances and at least $\radius+1$ on NO instances. Consequently, $\ISVP_\infty$ is $\NP$-complete for principal ideals of $R_n$, where $n$ ranges over powers of two.
\end{theorem}
\end{minipage}
\begin{proof}
The definition of $\radius$ gives $\radius+C>3\norm{\polyg}_1$. The bounds in~\eqref{eq:dominant-operator} with $M=\radius+C$ and Lemma~\ref{lem:gaussian-unit} therefore imply that $f\ne0$ and that every shortest nonzero vector of $(f)$ has the form $f\sigma$ with $\sigma\in\GU$. By Lemma~\ref{thm:principal-encoding}, $\theta_{\polyg}(\sigma)\le C$ for every $\sigma\in\GU$. The first minimum of $(f)$ is therefore at least $\radius$ by Lemma~\ref{thm:gaussian-conversion}. On a YES instance, some $\sigma\in\GU$ satisfies $\theta_{\polyg}(\sigma)=C$ and hence $\norm{f\sigma}_\infty=\radius$, so the first minimum equals $\radius$. On a NO instance, $\theta_{\polyg}(\sigma)\le C-1$ for every $\sigma\in\GU$ by integrality, so the first minimum is at least $\radius+1$.

The choice of $m$ gives $n=2m\le16(8s+1)$. Since $C=O(sW+B)$ and $\norm{\polyae}_1\le5^{m/2}$, we have $\norm{\polyg}_1=O(C5^{m/2})$. Thus $\radius=3\norm{\polyg}_1+1=O(C5^{m/2})$, and $\norm f_\infty\le\radius+C+\norm{\polyg}_1=O(C5^{m/2})$. Hence $\radius$ and each entry of $\matM_f$ have $O(s+\log(sW+B))=O(s+\beta+\log s)$ bits. The output basis $\matM_f$ has $n^2$ entries, so the output has $O(s^2(s+\beta+\log s))$ bits. The explicit formulas therefore give a deterministic polynomial-time construction. Membership in $\NP$ follows from Lemma~\ref{lem:integer-operations}.\qed
\end{proof}

\begin{samepage}
The proof of Theorem~\ref{thm:principal} also gives the following corollary. Comparing the norm of a shortest vector with $\radius$ decides the source \SSum\ instance. On a YES instance, the linear algebra in Section~\ref{subsec:ideal-module-svp} recovers $\sigma$ from a shortest vector $y=f\sigma$, and Lemma~\ref{thm:principal-encoding} recovers a solution to the \SSum\ instance in polynomial time.

\begin{corollary}\label{cor:principal-search}
Search-$\ISVP_\infty$ is $\NP$-hard for principal ideals of $R_n$ under polynomial-time Turing reductions, where $n$ ranges over powers of two.
\end{corollary}
\end{samepage}

\section[Transfer to cyclic rings]{Transfer to cyclic rings}\label{sec:cyclic}
In this section, we transfer the reductions in Sections~\ref{sec:module} and~\ref{sec:principal} from $R_n=\Z[X]/(X^n+1)$ to the cyclic polynomial rings $S_{2n}\coloneqq\Z[X]/(X^{2n}-1)$. We extend the scalar construction used for Euclidean CVP by Liu, Feng, and Pan~\cite{LFPCVP} to polynomial matrices. The construction preserves module rank while doubling the ring degree and the first minimum in the coefficient $\ell_\infty$-norm.

Each element of $S_{2n}$ has a unique representative $f=\sum_{\ell=0}^{2n-1}[f]_\ell X^\ell$ of degree less than $2n$. We identify it with the coefficient vector $([f]_0,\ldots,[f]_{2n-1})^{\mathsf T}\in\Z^{2n}$. Since $X^{2n}=1$, multiplication by $X$ cyclically permutes these coefficients. An ideal of $S_{2n}$ has \emph{full rank} if its coefficient lattice has rank $2n$.

Let $n\ge1$ be a power of two and $r\ge1$. Let $\matG=(\polyg_{ab})\in R_n^{r\times r}$ satisfy $\det\matG\ne0$, and represent its entries by polynomials of degree less than $n$. Set $F\coloneqq\max_{1\le a,b\le r}\norm{\polyg_{ab}}_\infty$ and $C\coloneqq2F+1$. Define
\begin{equation}\label{eq:cyclic-lift}
 \widehat{\matG}\coloneqq
 (1-X^n)\matG+C(1+X^n)\matI_r\in S_{2n}^{r\times r}.
\end{equation}
Over $\Q$, the Chinese remainder theorem gives
\[
 \Q\otimes_{\Z}S_{2n}
 \cong\Q[X]/(X^n+1)\times\Q[X]/(X^n-1).
\]
Reducing the matrix in~\eqref{eq:cyclic-lift} modulo $X^n+1$ gives $2\matG$, while reducing it modulo $X^n-1$ gives $2C\matI_r$.

\begin{samepage}
\begin{theorem}\label{thm:cyclic-transfer}
The matrix $\widehat{\matG}$ defined by~\eqref{eq:cyclic-lift} has the following properties.
\begin{enumerate}
\item {The module $\widehat{\matG}S_{2n}^r$ is free of rank $r$, and its coefficient lattice has rank $2rn$. For $r=1$, it is  a principal ideal of $S_{2n}$.}
\item {The first minima satisfy}
\begin{equation}\label{eq:cyclic-minimum}
 \lambda_{1,\infty}(\widehat{\matG}S_{2n}^r)
 =2\lambda_{1,\infty}(\matG R_n^r).
\end{equation}
\item {Every entry of $\widehat{\matG}$ has coefficients of absolute value at most $3F+1$.}
\end{enumerate}
\end{theorem}
\end{samepage}
\begin{proof}
For $\vecu\in S_{2n}^r$, write its canonical representative as $\vecu=\veca+X^n\vecb$ with $\veca,\vecb\in\Z[X]^r$ of degree less than $n$. Its images modulo $X^n+1$ and $X^n-1$ are $\veca-\vecb$ and $\veca+\vecb$, respectively. Let $\vecv=(v_1,\ldots,v_r)\in\Z[X]^r$ be the vector obtained by reducing each coordinate of $\matG(\veca-\vecb)$ modulo $X^n+1$ to degree less than $n$. Expanding~\eqref{eq:cyclic-lift} yields
\begin{equation}\label{eq:cyclic-block-identity}
 \begin{aligned}
 \widehat{\matG}\vecu
 &=(1-X^n)\matG(\veca-\vecb)
   +C(1+X^n)(\veca+\vecb)\\
 &=(1-X^n)\vecv+C(1+X^n)(\veca+\vecb)\\
 &=\bigl(\vecv+C(\veca+\vecb)\bigr)
   +X^n\bigl(-\vecv+C(\veca+\vecb)\bigr).
 \end{aligned}
\end{equation}
The first equality uses $X^{2n}=1$ in $S_{2n}$. The second uses the definition of $\vecv$ and $(1-X^n)(X^n+1)=0$.

Applying $\max\{|x+t|,|-x+t|\}=|x|+|t|$ to corresponding coefficients in the two blocks of~\eqref{eq:cyclic-block-identity} gives
\begin{equation}\label{eq:cyclic-norm-identity}
 \norm{\widehat{\matG}\vecu}_\infty
 =\max_{1\le j\le r,\ 0\le k<n}
 \bigl(|[v_j]_k|+C|[a_j+b_j]_k|\bigr).
\end{equation}
For (i), suppose $\widehat{\matG}\vecu=\mathbf0$. Equation~\eqref{eq:cyclic-norm-identity} gives $\veca+\vecb=\mathbf0$ and $\vecv=\mathbf0$. The definition of $\vecv$ and injectivity of $\matG$ give $\veca-\vecb=\mathbf0$, so $\veca=\vecb=\mathbf0$. Thus $\vecu=\mathbf0$, so the expanded integer matrix of $\widehat{\matG}$ is nonsingular, proving (i).

For (ii), set $\lambda\coloneqq\lambda_{1,\infty}(\matG R_n^r)\le F$ and consider any nonzero vector $\widehat{\matG}\vecu$. If $\veca+\vecb\ne\mathbf0$, equation~\eqref{eq:cyclic-norm-identity} gives $\norm{\widehat{\matG}\vecu}_\infty\ge C>2\lambda$. If $\vecb=-\veca$, set $\vecy\coloneqq\matG\veca\in R_n^r\setminus\{\mathbf0\}$ and represent each coordinate of $\vecy$ by a polynomial of degree less than $n$. Equation~\eqref{eq:cyclic-block-identity} gives $\norm{\widehat{\matG}\vecu}_\infty=2\norm{\vecy}_\infty\ge2\lambda$ and
\[
 \widehat{\matG}\vecu=2\vecy-X^n(2\vecy).
\]

Choose $\veca\in R_n^r$ with $\norm{\matG\veca}_\infty=\lambda$ and set $\vecu\coloneqq\veca-X^n\veca\in S_{2n}^r$ using the canonical representative of $\veca$. Equation~\eqref{eq:cyclic-block-identity} gives $\norm{\widehat{\matG}\vecu}_\infty=2\lambda$, proving~\eqref{eq:cyclic-minimum}.

For (iii), expanding $\widehat{\matG}=(\matG+C\matI_r)+X^n(-\matG+C\matI_r)$ shows that every coefficient has absolute value at most $F+C=3F+1$.\qed
\end{proof}

\begin{samepage}
\begin{corollary}\label{cor:cyclic-hardness}
For every fixed integer $r\ge1$, decision-$\SVP$ in the coefficient $\ell_\infty$-norm is $\NP$-complete for free modules of rank $r$ over $S_{2n}=\Z[X]/(X^{2n}-1)$ with $n$ a power of two. In particular, this holds for the following input classes.
\begin{enumerate}
\item {principal ideals of $S_{2n}$.}
\item {free modules of rank two over $S_{2n}$.}
\end{enumerate}
\end{corollary}
\end{samepage}
\begin{proof}
Take the output $(n,(f),\radius)$ of Theorem~\ref{thm:principal}. Set $\matG_1\coloneqq(f)$ and, for $r\ge2$, $\matG_r\coloneqq\operatorname{diag}(f,(\radius+1)\matI_{r-1})$. The added coordinates have norm at least $\radius+1$ when nonzero, so $\matG_rR_n^r$ has first minimum $\radius$ on YES instances and at least $\radius+1$ on NO instances. Theorem~\ref{thm:cyclic-transfer} transforms $\matG_r$ into a generator matrix $\widehat{\matG}_r$ over $S_{2n}$. Let $\mathcal M\coloneqq\widehat{\matG}_rS_{2n}^r$. We output the instance  $(2n,\mathcal M,2\radius)$. For $r=1$, we provide an integer basis of the ideal by taking the coefficient vectors of the cyclic shifts of the entry of $\widehat{\matG}_1$. By~\eqref{eq:cyclic-minimum}, the output has first minimum $2\radius$ on YES instances and at least $2\radius+2$ on NO instances.

For fixed $r$, the parameter bounds in Theorems~\ref{thm:principal} and~\ref{thm:cyclic-transfer}(iii), together with the explicit formula~\eqref{eq:cyclic-lift}, give a deterministic polynomial-time reduction. A nonzero coefficient vector of norm at most the input bound is a certificate of polynomial bit length. Integer linear algebra checks input validity, and Hermite normal form tests lattice membership in polynomial time~\cite{Cohen}. These decision problems therefore belong to $\NP$.\qed
\end{proof}

\begin{samepage}
Every nonzero vector in $\widehat{\matG}S_{2n}^r$ of norm less than $C$ has the form $2\vecy-X^n(2\vecy)$ with $\mathbf0\ne\vecy\in\matG R_n^r$. Taking its coefficients of degrees $0,\ldots,n-1$ in each coordinate and dividing by $2$ recovers $\vecy$ in polynomial time.

The proof of Corollary~\ref{cor:cyclic-hardness} also gives the following corollary. Comparing the norm of a shortest vector with $2\radius$ decides the source \SSum\ instance. On a YES instance, taking the first $n$ coefficients of its first coordinate and dividing by $2$ gives a shortest vector of $(f)$, from which Section~\ref{subsec:principal-correctness} recovers a solution in polynomial time.

\begin{corollary}\label{cor:cyclic-search}
Search-$\SVP$ in the coefficient $\ell_\infty$-norm is $\NP$-hard for principal ideals and free modules of any fixed rank $r\ge1$ over $S_{2n}$ under polynomial-time Turing reductions, where $n$ ranges over powers of two.
\end{corollary}
\end{samepage}

\section{Conclusion}\label{sec:conclusion}
We prove that decision-$\SVP$ in the coefficient $\ell_\infty$-norm is $\NP$-complete for principal ideals and  free modules of rank two over the power-of-two cyclotomic rings $R_n=\Z[X]/(X^n+1)$. Both results hold under deterministic polynomial-time reductions from the \SSum\ problem, with lattice dimension linear in the number of input integers.

We also establish $\NP$-completeness for principal ideals and free modules of every fixed positive rank over the power-of-two cyclic rings $S_{2n}=\Z[X]/(X^{2n}-1)$ by transferring the reductions from $R_n$. Over both ring families, the corresponding search problems are $\NP$-hard under polynomial-time Turing reductions.

% ============================================================
% AI disclosure
% ============================================================
\section*{Use of AI}

The reductions and proofs in this paper were found using OpenAI's
GPT-6.1 Sol (Ultra). The model initially produced a reduction from
NAE-3SAT, which gave a more complicated construction with a larger
lattice dimension and much larger constant factors. Working with the
model, we switched to Subset Sum and simplified the constructions
to obtain smaller constant factors and lattice dimensions linear in the number of input integers. We also used the model to generate the accompanying Lean
formalization. We reorganized and polished the manuscript, verified
all mathematical proofs and references ourselves, and take full
responsibility for the final paper.

% ============================================================
% References
% ============================================================
\ifnum\fullversion=1\relax
  \bibliographystyle{alpha-short}
\else
  \bibliographystyle{splncs03}
\fi

\bibliography{references}

\end{document}